\documentclass{JHEP3}
\pdfoutput=1
\let\ifpdf\relax

\usepackage{graphicx}
\usepackage{fancyvrb}
\usepackage{makeidx}
\usepackage{latexsym,amssymb,amsmath,makeidx}
\usepackage{ifpdf}

\newcommand{\beqa}{\begin{eqnarray}}
\newcommand{\eeqa}{\end{eqnarray}}
\newcommand{\beqs}{\begin{subequations}}
\newcommand{\eeqs}{\end{subequations}}
\newcommand{\beq}{\begin{equation}}
\newcommand{\eeq}{\end{equation}}
\newcommand{\bq}{\begin{equation}}
\newcommand{\eq}{\end{equation}}
\newcommand{\ba}{\begin{array}}
\newcommand{\ea}{\end{array}}
\newcommand{\beqn}{\begin{eqnarray}}
\newcommand{\eeqn}{\end{eqnarray}}
\newcommand{\be}{\begin{equation}}
\newcommand{\ee}{\end{equation}}

\newcommand{\mathsym}[1]{{}}

\def\nn{\nonumber}
\def\dis{\displaystyle}

\def\({\left(}
\def\){\right)}

\def\End{\end{document}}

\def\leqq{\leqslant}
\def\geqq{\geqslant}

\def\hf{\frac{1}{2}}
\def\td{\tilde}
\def\wtd{\widetilde}

\makeindex

\newcommand{\GeV}{\,\mathrm{GeV}}
\newcommand{\TeV}{\,\mathrm{TeV}}

\newcommand{\fb}{\,\mathrm{fb}}

\def\ifb{\text{fb}^{-1}}

\def\a{\alpha}

\def\ga{\gamma}

\def\mA{\mathcal{A}}

\def\mI{\mathcal{I}}

\def\mL{\mathcal{L}}

\def\mN{\mathcal{N}}
\def\mO{\mathcal{O}}

\def\tanb{\tan\!\beta}
\def\tant{\tan\theta}
\def\Nu{\nu_4^{}}

\def\Nuu{\nu_4^{}{\nu}_4^{}}
\def\NN{{\cal N}_4^{}}
\def\NNN{{\cal N}_4^{}{\cal N}_4^{}}

\def\T{t_4^{}}
\def\B{b_4^{}}

\def\MA{M_A^{}}

\def\Mh{M_h^{}}
\def\Mhh{M_h^2}
\def\mf{m_f^{}}

\title{LHC Signatures of Two-Higgs-Doublets\\ with Fourth Family}

\author{Ning Chen\,$^{a}$ ~and~ Hong-Jian He\,$^{a,b,c}$\\
        $^a$\,Institute of Modern Physics and Center for High Energy Physics,\\
             ~\,Tsinghua University, Beijing 100084, China.\\
        $^b$\,Center for High Energy Physics, Peking University, Beijing 100871, China.\\
        $^c$\,Kavli Institute for Theoretical Physics China, CAS, Beijing 100190, China. \\

       E-mail: \email{hep\_nchen@tsinghua.edu.cn, hjhe@tsinghua.edu.cn}
       }

\abstract{On-going Higgs searches in the light mass window are of vital importance for testing the Higgs mechanism and probing new physics beyond the standard model (SM). The latest ATLAS and CMS
searches for the SM Higgs boson at the LHC\,(7\,TeV) found some intriguing excesses of events
in the \,$\gamma\gamma/VV^*$\, channels ($V=Z,W$) around the mass-range of \,$124-126$\,GeV.
We explore a possible explanation of the $\gamma\gamma$ and $VV^*$ signals from the light
CP-odd Higgs $A^0$ or CP-even Higgs $h^0$ from the general two-Higgs-doublet model with
fourth-family fermions. We demonstrate that {\it by including invisible decays of the
Higgs boson $A^0$ or $h^0$ to fourth-family neutrinos,} the predicted $\gamma\gamma$
and $VV^*$ signals can explain the observed new signatures at the LHC, and will be
further probed by the forthcoming LHC runs in 2012.}

\keywords{Higgs Physics, Beyond Standard Model 
\\[2mm]
JHEP (2012), in Press, [\,arXiv:1202.3072\,]}

\begin{document}


\maketitle

\tableofcontents

\setcounter{page}{2}

\section{\hspace*{-1mm}Introduction}
\label{sec:1}
\vspace*{2mm}

The LHC searches for Higgs boson(s) in the light mass window
have vital importance for testing the Higgs mechanism \cite{Higgs}
and probing new physics beyond the SM. The most recent results from the LHC\,(7\,TeV) 
have constrained the light Higgs boson of the standard model (SM)
into the mass-range \,(115.5\,GeV,\, 131\,GeV)
by ATLAS\,\cite{Atlas2011-12} and
\,(115\,GeV,\, 128\,GeV) by CMS\,\cite{CMS2011-12}, at $95\%$\,C.L.\footnote{From the
latest updates at the Moriond conference \cite{Mori},\,
ATLAS further confined the allowed light SM Higgs mass ranges into
\,(117.5\,GeV,\, 118.5\,GeV) and \,(122.5\,GeV,\, 129\,GeV) at $95\%$\,C.L.,
while CMS gave the improved Higgs mass limits of \,(114.4\,GeV,\, 127.5\,GeV).}
In particular, the ATLAS observed
an intriguing excess of events for
a Higgs boson with mass close to $\,m_{h}^{}=126\GeV$ \cite{Atlas2011-12}.
The three most sensitive channels in this mass range,
$\,h^0\to\gamma\gamma$,\, $h^0\to ZZ^*\to\ell^+\ell^-\ell^+\ell^-$,\, and \,$h^0\to WW^*\to\ell^+\nu\ell^-\bar\nu$,\,
contribute to the excess with local significances of
$\,2.8\sigma$, \,$2.1\sigma$,\, and $\,1.4\sigma$,\, respectively.
If this would be confirmed by the upcoming LHC data in 2012, a Higgs boson of mass
around 126\,GeV does call for new physics beyond the SM due to the
vacuum instability \cite{vacS}. Furthermore, the observed $\,2.8\sigma$
excess in the $\gamma\gamma$ channel by ATLAS is also
higher than the expected signals of the pure SM Higgs boson (with the same mass) by a factor-2 \cite{Atlas2011-12}, which again points to new physics.

In this work, we investigate a simple SM-extension as the new physics ---  the generic two-Higgs-doublet model (2HDM) with fourth-family SM fermions (4F2HDM). It contains the minimal extension in the SM Higgs sector with one more doublet and in the SM fermion sector with one more family. With such a truly simple addition, we study distinct new signatures of the light CP-odd Higgs $A^0$ or CP-even Higgs $h^0$ at the LHC, and analyze the implications for the latest ATLAS and CMS Higgs searches \cite{Atlas2011-12}\cite{CMS2011-12}.
We consider the 4F2HDM in both type-I and type-II, with CP-conserving Higgs potential. Such 2HDMs contain four physical Higgs states
$(h^0,\,H^0,\,A^0,\,H^\pm)$ with masses \,$(\Mh,\, M_H^{},\, \MA,\, M_{\pm}^{})$.\,
Due to the additional contributions from heavy fourth-family quarks $(\T,\,\B)$,
the gluon-fusion production cross sections of $\,gg\to h^0,A^0\,$
at the LHC are generally much enhanced relative to $\,gg\to h^0\,$ in the SM,
and thus may be easily excluded by the current LHC data.
In the present study, we demonstrate that the invisible Higgs decays
into the light fourth-family neutrinos, $\,h^0,A^0\to\Nuu,\NNN$,\,
can become the major channel, and play a key role to properly suppress
$\,h^0,A^0\to\gamma\gamma$\, rates for the consistency
with the existing LHC data.
 Especially, we show that such a light Higgs boson $h^0$ or $A^0$
 with mass around $124-126$\,GeV can nicely explain the observed
 event excesses by ATLAS\,\cite{Atlas2011-12} and CMS\,\cite{CMS2011-12}.


\vspace*{4mm}
\section{\hspace*{-1mm}Signals of CP-Odd ${\mathbf A}^{\!\mathbf 0}$
                       in 4F2HDM with Invisible Decays}
\vspace*{2mm}
\label{sec:2}

We start with the analysis of CP-odd Higgs boson $A^0$.\,
The general 2HDM allows $A^0$ to be the lightest Higgs boson for
proper parameter space of the Higgs potential, unlike the minimal supersymmetric SM (MSSM) where the lightest Higgs boson is always $h^0$.\,
An explicit realization of such 2HDMs is given
by the dynamical top-seesaw model \cite{topseesaw-rev},
where the light mass of the composite pseudo-scalar $A^0$
is induced by the topcolor instanton effect \cite{He:2001fz}
and thus can naturally serve as the lightest state in the Higgs spectrum.
Since $A^0$ has no cubic gauge couplings at tree-level,
it mainly decays into the SM fermion pairs
and the $\,gg/ \gamma\gamma\,$ final states (via triangular fermion-loops). So the decay channel $\,A^0\to \gamma\gamma\,$ could be important for detecting such a light $A^0$ boson at the LHC. However, it was recently found\,\cite{Gunion:2011ww, Burdman:2011ki} that a light $A^0$ in the presence of fourth-family is excluded due to the enhanced cross section and unsuppressed decay branching ratio of $\,A^0\to \gamma\gamma\,$. We note that this exclusion holds only in certain parameter region. In the following, we will include the invisible decays $\,A^0\to\Nuu,\NNN$\, for the fourth-family neutrinos being lighter than half of $\,\MA$,\, and study the distinct new LHC signatures of the $A^0$ Higgs boson.

The Higgs potential of the general 2HDM contains two characteristic
input parameters, the $\,\tan\beta\equiv v_1^{}/v_2^{}$\, as the ratio of two Higgs vacuum expectation values (VEVs), and
the mixing angle $\,\alpha$\, from diagonalizing the mass-matrix of
neutral Higgs bosons \,$(h^0,\,H^0)$\,.\,
It was shown \cite{He:2001tp} that such 2HDM with fourth-family fermions
is consistent with the electroweak precision constraints. 
Ref.\,\cite{BarShalom:2011zj} also found that within broad parameter regions,
the 4F2HDM can satisfy the $\,\bar{B}\to X_s\gamma$\, and $\,B_q^{}-\bar{B}_q^{}$\,
mixing constraints.  For the present study, we focus on two types of CP-conserving 2HDMs 
without tree-level flavor-changing neutral currents (FCNC) \cite{Branco:2011iw}, 
the type-I and type-II 2HDMs including the fourth-family. 
By definition, the type-I 2HDM assigns the first Higgs doublet 
$\Phi_1^{}$ (with VEV $v_1^{}$) to couple with all fermions via Yukawa interactions and
generate their masses, but the second Higgs doublet
$\Phi_2^{}$ (with VEV $v_2^{}$) does not. The type-II 2HDM has
$\Phi_1^{}$ couple to all up-type fermions and $\Phi_2^{}$ to all down-type fermions.
The most general Yukawa interactions for
the pseudo-scalar $A^0$ in the 4F2HDM can be expressed as,
\beqn
\mL_{\rm Yukawa} \,&=&\,
-\sum_f\frac{\mf}{v}\xi_A^f\overline{f}i\gamma_5^{} f A^0 \,,
\eeqn
where the couplings $\,\xi_A^f\,$ in the 4F2HDM-I and -II are summarized
in Table\,\ref{tab:A-Yukawa}.
\begin{table}[h]
\begin{center}
\begin{tabular}{c||c|c}
\hline\hline
   & ~4F2HDM-I~ & ~4F2HDM-II~     \\
   \hline
 $~\xi_A^u~$  & $\cot\beta$        & $\cot\beta$   \\
 $\xi_A^d$    & $-\cot\beta$~~    & $\tan\beta$   \\
 $\xi_A^{\nu}$  & $\cot\beta$      & $\cot\beta$   \\
 $\xi_A^{\ell}$  & $-\cot\beta$~~ & $\tan\beta$  \\
\hline\hline
\end{tabular}
\caption{Yukawa couplings of the CP-odd Higgs boson $A^0$
         in the 4F2HDM-I and 4F2HDM-II.}
\label{tab:A-Yukawa}
\end{center}
\end{table}

\begin{figure}
\centering
\includegraphics[width=11cm,height=8.5cm]{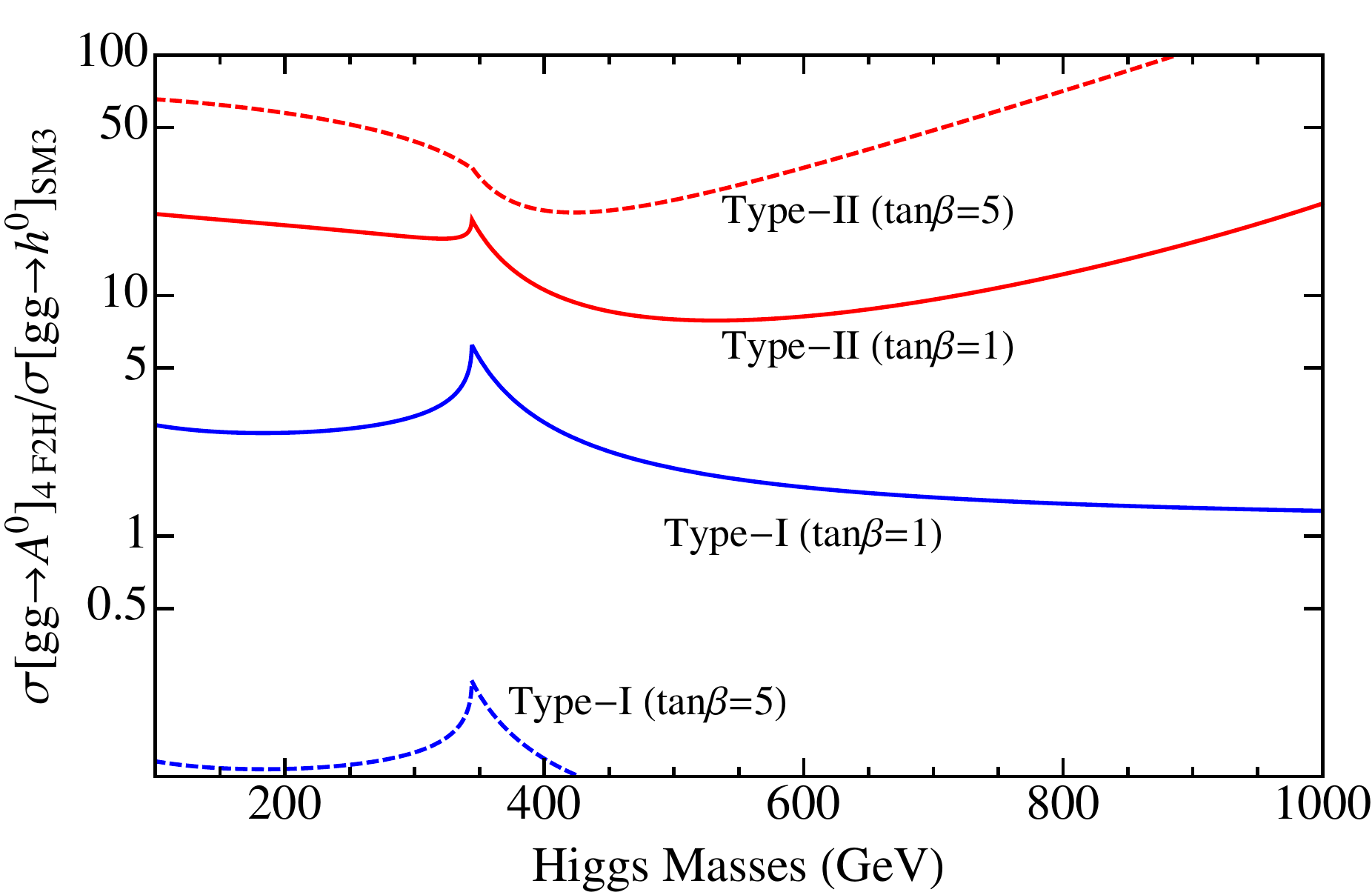}
\vspace*{-2mm}
\caption{Ratio of $\,\sigma[gg\!\to\! A^0]_{\rm 4F2H}^{}/\sigma[gg\!\to\! h^0]_{\rm SM3}^{}\,$ for $\,\tanb =1$\, (solid lines) and $\,\tanb=5$\, (dashed lines).}
\label{fig:ggA}
\end{figure}

The major production channel of $A^0$ at the LHC is the gluon-fusion process, and its cross section differs from that of the SM Higgs boson (coupled to three families of SM fermions) through the ratio,
\beqn
\label{eq:ggA-enhance}
\frac{\,\sigma[gg\!\to\! A^0]_{\rm 4F2H}^{}\,}{\sigma[gg\!\to\! h^0]_{\rm SM3}^{}}
~=~ \frac{\,\left|\sum_{Q=\T, \B, t}\xi_A^Q
             \mI_A^{}(\tau_Q^{})\right|^2}{\left|\mI_S^{}(\tau_t)\right|^2}\,
. ~~~~~~
\eeqn
Here the form factors for the CP-even and CP-odd Higgs bosons
take the forms \cite{Spira:1995rr},
\beqn
&& \mI_S^{}(\tau) \,=\, \frac{1}{\tau^2}\left[\tau+(\tau \!-\! 1)f(\tau)\right],
~~~~
\mI_A^{}(\tau) \,=\, \frac{1}{\tau}f(\tau) \,,~~~~~~~
\\[3mm]
&& f(\tau) \,=\, \begin{cases}
~ \arcsin^2\!\sqrt{\tau}\,,  \qquad & \tau\leqq 1\,,
\\[2.5mm]
\dis ~ -\frac{1}{4}\!
 \left[\ln\frac{1+\sqrt{1-\tau^{-1}}}{1-\sqrt{1-\tau^{-1}}}-i\pi\right]^2\!,~~
\qquad & \tau>1 \,,
\end{cases}
\eeqn
with $\,\tau_f^{}\equiv M_{h, A}^2/(4m_f^2)$.\,
Notice that the ratio of the on-shell production cross sections (\ref{eq:ggA-enhance})
is clearly independent of the center-of-mass energy of the LHC.
This is also true for the ratio of the corresponding signal event numbers,
as the integrated luminosity is the same for both cross sections.
Hence our predicted ratio of signals for either production cross sections or number of events should also apply to the forthcoming LHC runs with higher collision energies and/or higher luminosities \cite{LHC2012}. Then, we compute the ratio (\ref{eq:ggA-enhance}) for the inputs $\,\tanb =1\,$ and $\,\tanb =5\,$ in Fig.\,\ref{fig:ggA}. For larger $\tanb$ vlaues, the ratio (\ref{eq:ggA-enhance}) for 4F2HDM-II receives an enhancement from fourth-family quark $\,\B\,$\,
$\sim\tan^2\!\beta\,|\mI_A^{}(\tau_{\B}^{})/\mI_S^{}(\tau_t^{})|^2$\,.\,
Thus the production cross section $\,\sigma[gg\to A^0]_{\rm 4F2H}$\,
is more enhanced for large $\,\tan^2\!\beta\,$ relative to that of the SM Higgs boson. On the other hand, all type-I Yukawa couplings are controlled by an overall factor $\,\cot\!\beta\,$  as shown in Table\,\ref{tab:A-Yukawa}. So the fourth-family quarks give contributions proportional to a uniform factor $\sim\cot^2\!\beta\,|\mI_A^{}(\tau_{Q}^{})/\mI_S^{}(\tau_t^{})|^2$\,.\, Obviously, the fourth-family corrections to the gluon-fusion cross section is enhanced by $\,\tan^2\!\beta\,$ in 4F2HDM-II while suppressed by $\,\cot^2\!\beta\,$ in 4F2HDM-I for $\,\tanb <1\,$.\, Fig.\,\ref{fig:ggA} shows that for $\,\tanb \geqq 1\,$,\,
the $A^0$ production is always enhanced in 4F2HDM-II, and the
enhancement factor is about $\,{\cal O}(20-60)\,$ for $\,\tanb=1-5\,$ in the mass-range $\,M_A<300\,$GeV.\,
In contrast, the $A^0$ production in 4F2HDM-I is moderately enhanced by a factor of $\,{\cal O}(3-10)\,$ for $\,\tanb=1\,$ and $\,M_A<300\,$GeV,\,
which is much lower than that of 4F2HDM-II with the same $\,\tanb=1\,$.\,
Due the opposite signs between the up-type and the down-type Yukawa couplings $\,\xi_A^u\,$ and $\,\xi_A^d\,$ of 4F2HDM-I (Table\,\ref{tab:A-Yukawa}),
a cancellation appears between their contributions to the ratio (\ref{eq:ggA-enhance}).
This cancellation becomes maximal when the two heavy quarks $(\T,\,\B)$ are degenerate.
So the cross sections in (\ref{eq:ggA-enhance}) are dominated by the
third-family top-quark-loop and the inequality
$\,\mI_A^{}(\tau_t^{})>\mI_S^{}(\tau_t^{})\,$
for a given Higgs mass determines the final enhancement of the ratio (\ref{eq:ggA-enhance})
for $\,\tanb=1\,$,\, as shown in Fig.\,\ref{fig:ggA} for 4F2HDM-I.
We also see that for a larger $\,\tanb\,$,\, such as $\,\tanb=5\,$,\,
the $A^0$ production in the 4F2HDM-I is suppressed by about a factor-10
relative to that of the $h^0$ in the SM3.

In general, the type-II Higgs sector is more nontrivial and interesting than the type-I, it is also well motivated for the fermion mass generations. In the natural parameter-space of $\,\tanb\gtrsim 1\,$,\,
it is very challenging to make the 4F2HDM-II safe from the LHC constraints as noted before\,\cite{Gunion:2011ww}\cite{Burdman:2011ki}. We have to sufficiently reduce the signals by suppressing the
relevant decay branching fractions of $\,A^0\,$.\,  For this purpose, we propose a new resolution by exploring the invisible decays of $\,A^0\,$ into light fourth-family neutrinos, $\,A^0\to \Nuu\,/\mN_4\mN_4$\,.\,

Generally, the fourth-family neutrinos \,$(\td{\nu}_4^{},\,\wtd{\cal N}_4^{})$\,
have both Dirac and Majorana mass-terms which form the seesaw mass-matrix,
\beqn
\left(
\begin{array}{cc}
0 ~&~ m_D^{} \\[1.5mm]
m_D^{} ~&~ M_N^{}
\end{array} \right).
\eeqn
After the diagonalization into mass-eigenbasis \,$({\Nu},\,{\NN})$\,,\,
their mass-eigenvalues are determined by the two mass-parameters
$\,m_D^{}\,$ and $\,M_N^{}\,$,
\beqn
&& M_{\Nu, \mN_4^{}}^{} ~=~ \sqrt{\frac{1}{4}{M_N^2}+m_D^2} \mp \hf{M_N^{}} \,,
\eeqn
with the mixing angle defined as,
\beqn
&& \tan\theta ~=\, \frac{\,M_{\nu_4}\,}{m_D^{}} \,=\, \frac{m_D^{}}{\,M_{\mN_4}\,}
\,=\, \sqrt{\frac{\,M_{\nu_4}\,}{\,M_{\mN_4}\,}} \,,
\eeqn
where $\,\tant \leqq 1\,$ must hold due to $\,M_{\nu_4}\leqq M_{\mN_4}\,$.\,
The case of $\,\tan\theta =1\,$ corresponds to $\,M_N^{}=0\,$,
leading to two degenerate states of pure Dirac neutrinos.
The limit of $\,\tan\theta =0$\, is unphysical
since it gives $\,M_{\nu_4}=0\,$.\, The LEP precision data on invisible $Z$ decays constrain $\,M_{\nu_4}\gtrsim \hf m_Z^{}\,$,\,
while the naturalness requires Yukawa couplings to be of $\,\mO(1)\,$
and thus the Dirac mass $\,m_D^{}=\mO(100-500)$GeV.\,
Hence, our parameter space for the mixing angle $\,\theta$\,
is confined into the range of $\,0.1\lesssim\tan\theta\leqq 1$\,.\,
The fourth-family neutrino $\Nu$ can be stable on the collider lifetime,
and the current experimental lower limits on stable neutral heavy lepton mass is
as low as $39.5$\,GeV at 95\%\,C.L., as inferred from the invisible $Z$ width \cite{PDG-2010,LEP-DEL92}.
Taking into account of the mixing between two Majorana neutrinos $\Nu$ and $\mN_4$,
this bound may be further reduced to $33.5$\,GeV \cite{Carpenter:2011wb}.
Such light fourth-family neutrinos will open up new invisible decay channels for both $\,A^0\,$ and $\,h^0$\,.\, The lower limit on the mass of fourth-family charged lepton $\ell_4^{}$ is about $100$\,GeV,
as given by the LEP-II direct searches \cite{LEP-L3-2001}.
These limits show that the fourth-family neutrinos $(\Nu,\,\NN)$ and
leptons ($\ell_4^{}$) can be much lighter than the fourth-family quarks $(\T,\,\B)$.
For short-lived \,$(\T,\,\B )$\, with prompt decays of $\,\T\!\to\! bW\,$
and $\,\B\!\to\!tW$,\, the current searches at the LHC\,(7\,TeV)
places the following lower bounds (95\%\,C.L.),
$\,M_{\T}>552$\,GeV from the CMS with $\mL=4.7\, \ifb$ \cite{CMS-t'-2012-1-31} or $\,M_{\T}>404$\,GeV from the ATLAS with $\mL=1.04\, \ifb$ \cite{ATLAS-t'-2012-2-14}, and $\,M_{\B}>495$\,GeV \cite{LHC-Mt'-Mb'}.
Meanwhile, the latest analysis from the Tevatron searches \cite{Tevatron-bound_4Gquark}
using an integrated luminosity of $\,12\fb^{-1}$\, for both CDF and D0 places the lower mass limits,  $\,M_{\T}>358$\,GeV\, and \,$M_{\B}>372$\,GeV\, at 95\%\,C.L.
For illustration in the following analysis,
we will uniformly take a sample input of fourth-family fermion masses,
$\,(M_{\T}^{},\, M_{\B}^{},\,  M_{\ell_4^{}}^{}, \, M_{\nu_4^{}}^{})
=(600,\, 600,\, 300, \, 50)$\,GeV,\, unless specified otherwise.

\begin{figure}
\centering
\includegraphics[width=12cm,height=9.5cm]{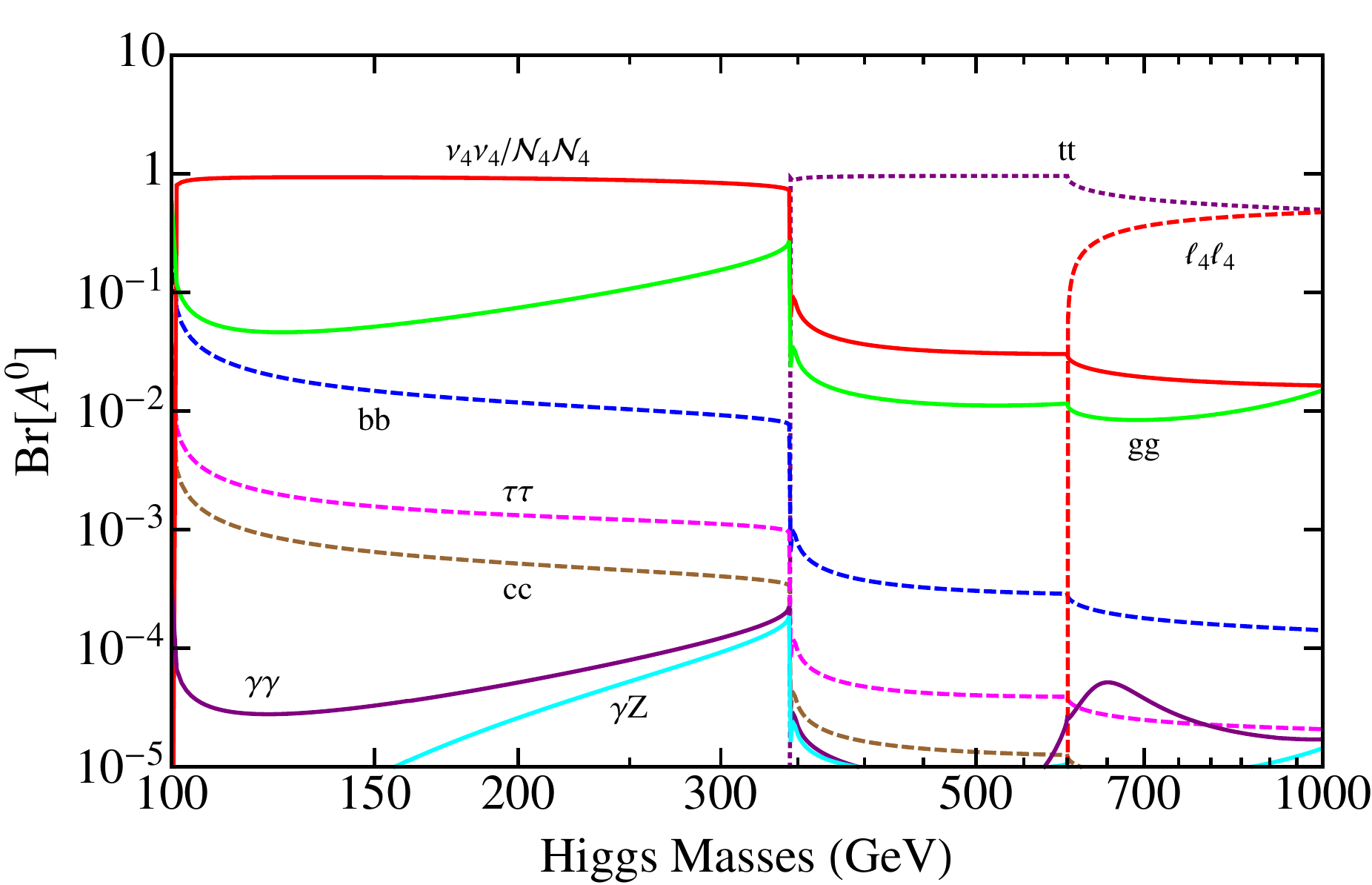}
\vspace*{-1mm}
\caption{Decay branching ratios of the CP-odd Higgs boson $\,A^0$\,
as functions of its mass $\,M_{A}^{}$\, for the 4F2HDM-II.
The other input parameters are fixed as, $\,(\tanb,\,\tan\theta) = (1,\,1)$\,.}
\label{fig:BRA-mA}
\end{figure}

For the present analysis, we will systematically explore the new decay channels of $\,A^0$\, in the 4F2HDM, $\,A^0\to \Nuu, \NNN$,\,
as well as $\,A^0\to \ell_4^{}\bar\ell_4^{}$,\,
when $A^0$ is heavier than twice of $\Nu$ ($\NN$) or $\ell_4^{}$.\, The invisible decay widths of $\,A^0\,$ are computed at the tree-level,
\beqa
\label{eq:A-nu4nu4}
\Gamma(A^0\!\to\!\Nuu) &\,=\,&
\frac{M_{\nu_4^{}}^2 M_{A}^{}|\xi_A^{\nu}|^2}{\,4\pi v^2(1\!+\!\tan^2\!\theta)^2\,}
\left(1\!-\!\frac{4M_{\nu_4}^2}{M_A^2} \right)^{\!\!\hf} ,
\\
\Gamma(A^0\!\to\!\mN_4\mN_4) &\,=\,&
\frac{M_{\mN_4^{}}^2 M_{A}^{}|\xi_A^{\nu}|^2}{\,4\pi v^2(1\!+\!\tan^2\!\theta)^2\,}
\left(1\!-\!\frac{4M_{\mN_4}^2}{M_A^2} \right)^{\!\!\hf} ,
\label{eq:A-N4N4}
\eeqa
where the second channel (\ref{eq:A-N4N4}) is open when
$\,M_{\mN_4}<\hf M_A$.\,
The fourth-family fermions also contribute to the loop-induced decay widths for $\,A^0\to gg,\,\gamma\gamma,\,\gamma Z\,$ as follows,
\beqn
\Gamma(A^0\!\to\! gg)_{\rm 4F2H}^{}
&\,=\,&
\frac{\alpha_s^2 M_{A}^3}{\,32\pi^3 v^2\,}\Big|
\!\sum_{Q=\T, \B, t}\!\!\!\xi_A^Q\mI_A^{}(\tau_{Q}^{})\Big|^2 ,
\label{eq:width-A-gg}
\\[2mm]
\Gamma(A^0\!\to\!\gamma\gamma)_{\rm 4F2H}^{}
&\,=\,&
\frac{\alpha^2 M_{A}^3}{64\pi^3 v^2}
\Big|\!\sum_{f}\! N_{c}^f e_f^2\xi_A^f\mI_A^{}(\tau_f^{}) \Big|^2,~~~~~~~~~
\label{eq:width-A-2photon}
\eeqn
\beqn
\Gamma(A^0\!\to\!\gamma Z)_{\rm 4F2H}^{}
&\,=\,&
\frac{\,\alpha M_{A}^3 m_W^2\,}{32\pi^4 v^4}\(1-\frac{m_Z^2}{M_{A}^2}\)^{\!\!3}
\Big|\!\sum_{f}\xi_A^f N_{c}^f
   \frac{e_f^{}c_f^{}}{c_W^{}}\wtd{\cal I}_A^{}(\tau_f^{}, \lambda_f^{}) \Big|^2
   ,~~~~~
\label{eq:width-A-Zgamma}
\eeqn
where $\,e_f^{}\,$ and $N_c^f$ denote the electric charge
and color-factor for each fermion species.
Besides, $\,c_f^{}\equiv 2T_{3f}-4e_f^{} s_W^2$,\, and \,$(s_W^{},\,c_W^{})\equiv (\sin\theta_W^{},\,\cos\theta_W^{})$\, with $\theta_W^{}$ being the weak mixing angle.
All decay widths in our analysis are computed by including the
relevant NLO QCD corrections as in Ref.\,\cite{Djouadi:1997yw}.
The form factor $\,\wtd{\cal I}_A^{}$\, in (\ref{eq:width-A-Zgamma}) is given by
\beqn
\wtd{\cal I}_A^{}(\tau_f^{}, \lambda_f^{})
~=\,\frac{\,f(\tau_f^{})-f(\lambda_f^{})\,}{2(\tau_f^{}-\lambda_f^{})} \,,
\eeqn
with $\,\lambda_f^{}\equiv m_Z^2/(4m_f^2)$\,.\,
Since the form factors $\,\mI_A(\tau_f^{})$\, and
$\,\wtd{\cal I}_A^{}(\tau_f^{}, \lambda_f^{})$ are positive
for the fermionic contributions, all three decay widths in (\ref{eq:width-A-gg})-(\ref{eq:width-A-Zgamma}) are larger than
$\Gamma(\,h^0\to gg,\,\gamma\gamma,\,\gamma Z\,)$ in the SM3.
Including the new invisible decay channels of $\,A^0\to\Nuu/\mN_4^{}\mN_4^{}$\,
with decay rates in (\ref{eq:A-nu4nu4})-(\ref{eq:A-N4N4})\,,\,
it is possible to suppress all SM decay branching fractions
for the low-mass range of $A^0$.\, In Fig.\,\ref{fig:BRA-mA},
we show the $A^0$ decay branching ratios in a wide mass-range of
$\,M_A^{}=100-1000$\,GeV\, for the 4F2HDM-II.
We take the sample inputs of \,$(\tanb,\,\tan\theta) = (1,\,1)$\,,\,
where $\,\tan\theta = 1$\, corresponds to the case of $\,\Nu$\, being pure Dirac neutrino.
Fig.\,\ref{fig:BRA-mA} shows that the invisible decay $\,A^0\to\Nuu, \mN_4\mN_4$\,
can dominate over all other channels for $\,M_{A} < 2m_t^{}$\,,
while the $\,t \bar t\,$ and $\,\ell_4^{}\bar{\ell}_4^{}$\, channels
become dominant for $\,M_{A} > 2m_t^{}$\,.\,
In particular, the diphoton channel $\,A^0\to \ga\ga$\,
can be suppressed by a factor of $\,{\cal O}(10-50)\,$ for $\,M_{A} < 2m_t^{}$\,,\,
as compared to the diphoton branching fraction of the SM Higgs boson in the same mass range.
Combining this with the enhanced cross sections in Fig.\,\ref{fig:ggA},
we see that the new invisible decays of $A^0$ play a key role
to bring down the $A^0$ signals for being consistent with the current LHC searches.
Moreover, $A^0$ has vanishing cubic couplings with gauge bosons and thus no $VV^*$
final states will be produced. It is clear that the current limits on the mass-range of
$A^0$ will be much weaker than that of the conventional SM Higgs boson
(mentioned at the beginning of Sec.\,\ref{sec:1}).

\begin{figure}
\centering
\includegraphics[width=10cm,height=8cm]{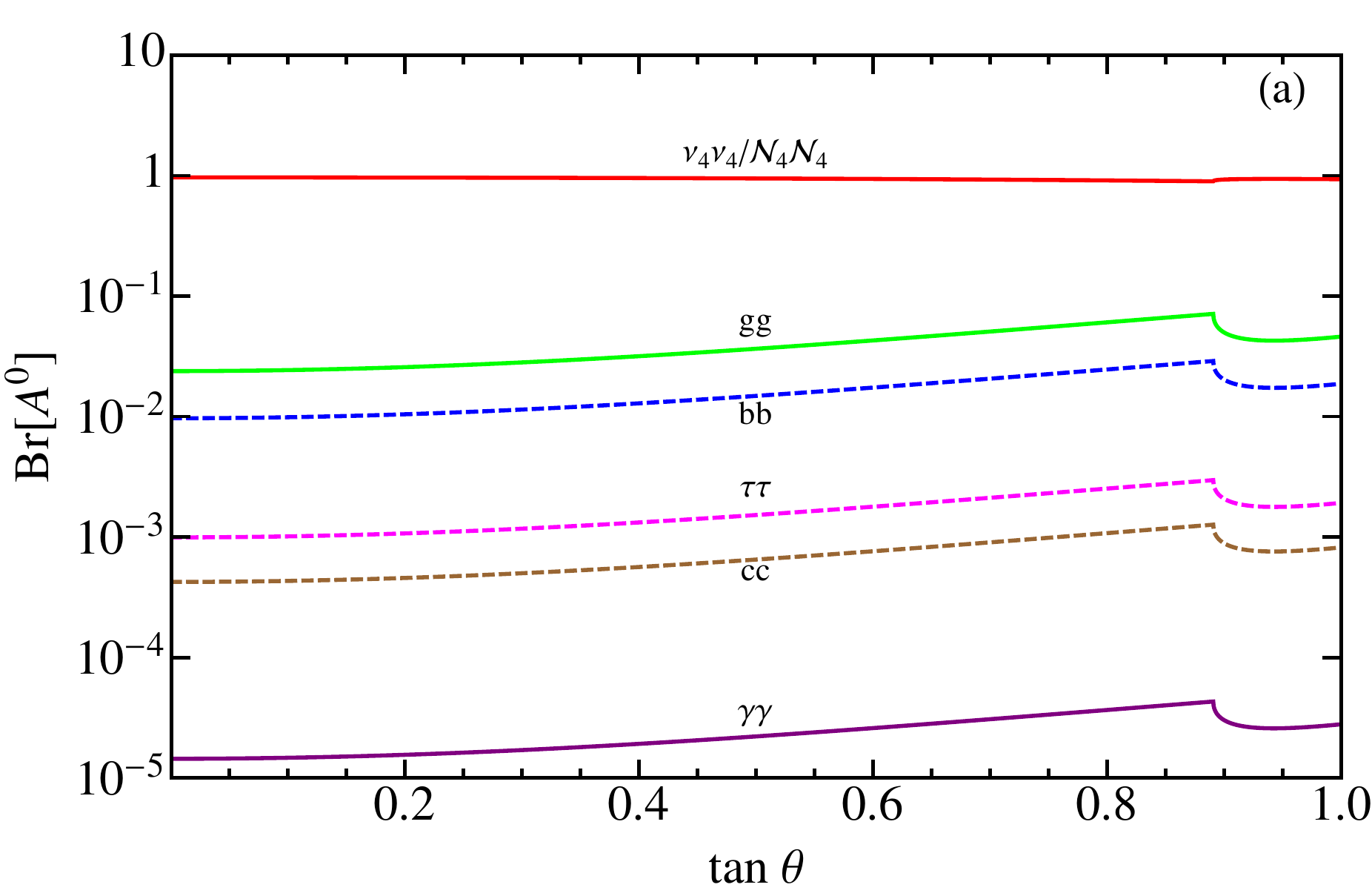}
\\[2mm]
\includegraphics[width=10cm,height=8cm]{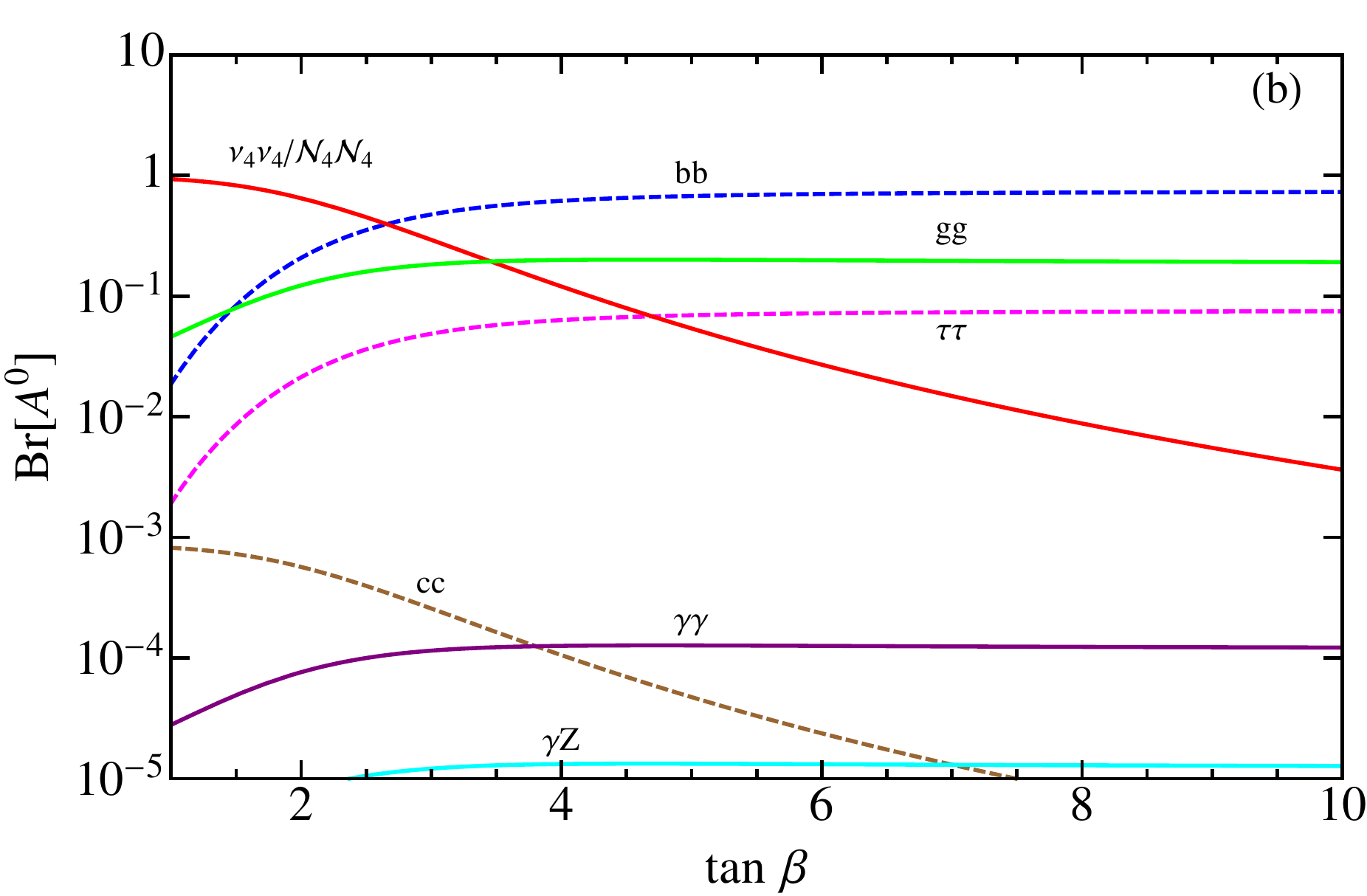}
\vspace*{-1mm}
\caption{Decay branching ratios of the CP-odd Higgs boson $\,A^0\,$
with mass $M_{A}=126$\,GeV\, for the 4F2HDM-II.
Plot-(a): Br$[A^0]$ as a function of the fourth-family neutrino mixing angle
$\,\tant\,$,\, with fixed $\,\tanb = 1$\,.
Plot-(b): Br$[A^0]$ as a function of $\,\tanb$\,,\, with fixed $\,\tant=1$\,.\,}
\label{fig:BRA-tanB-tanT}
\end{figure}

Motivated by the latest ATLAS data\,\cite{Atlas2011-12},
we focus on the case of a light $A^0$ with mass $\,M_{A}^{}=126$\,GeV. The invisible decay mode is kinematically allowed for
$\,39.5\GeV \lesssim M_{\nu_4} \lesssim 63\GeV$,\,
where the lower limit comes from the LEP constraints \cite{PDG-2010,LEP-DEL92}. For the invisible decay rates (\ref{eq:A-nu4nu4})-(\ref{eq:A-N4N4}), we have included both Dirac and Majorana neutrino masses. In the pure Dirac-mass limit $\,\tan\theta = 1\,$,\,
the two fourth-family neutrinos become degenerate and thus
their decay rates are equal, $\,\Gamma(A^0\to\Nu\bar{\nu}_4^{})= \Gamma(A^0\to\mN_4\bar\mN_4)$\,.

\begin{figure}
\centering
\includegraphics[width=10cm,height=8cm]{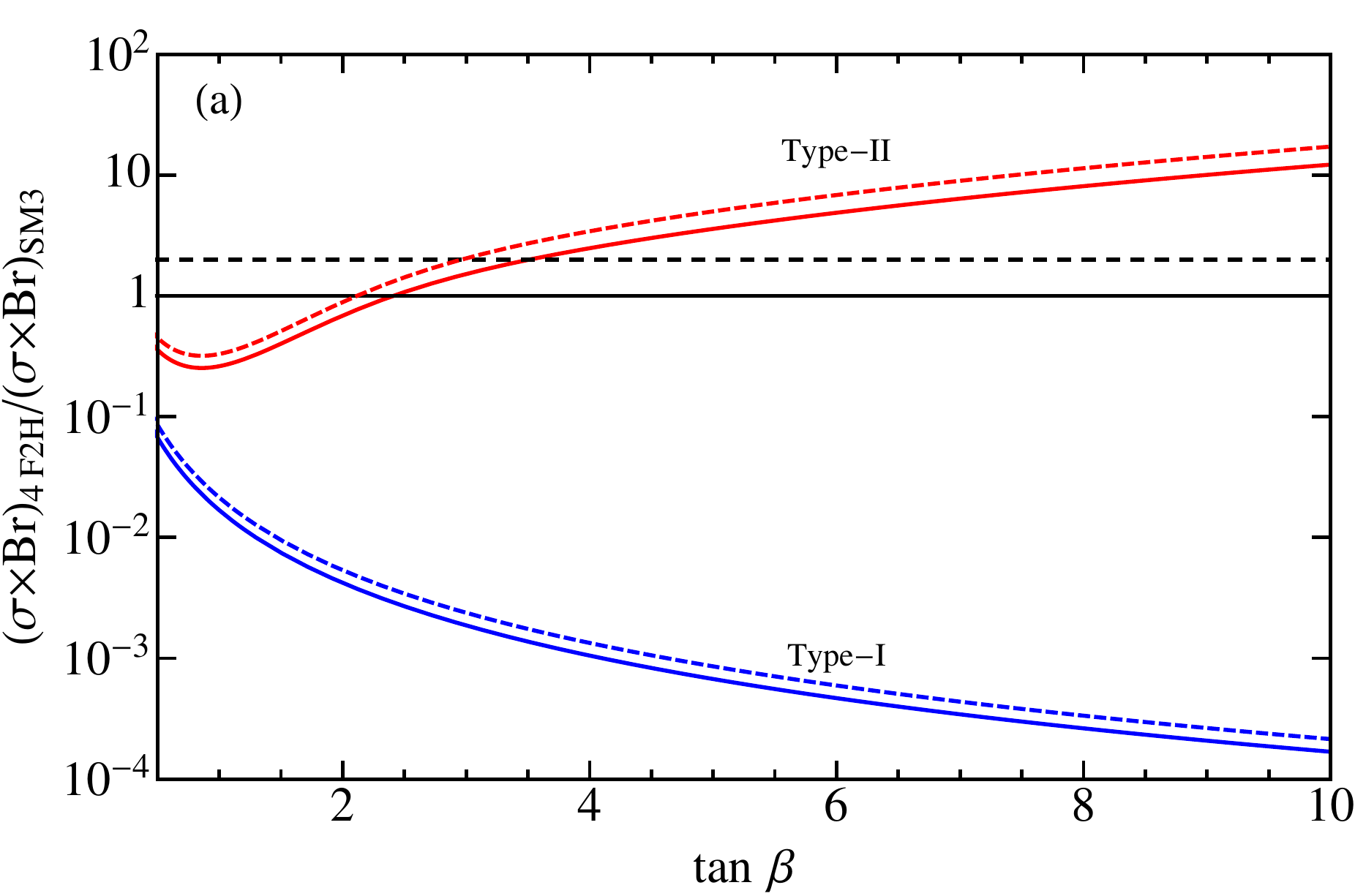}
\\[2mm]
\includegraphics[width=10cm,height=8cm]{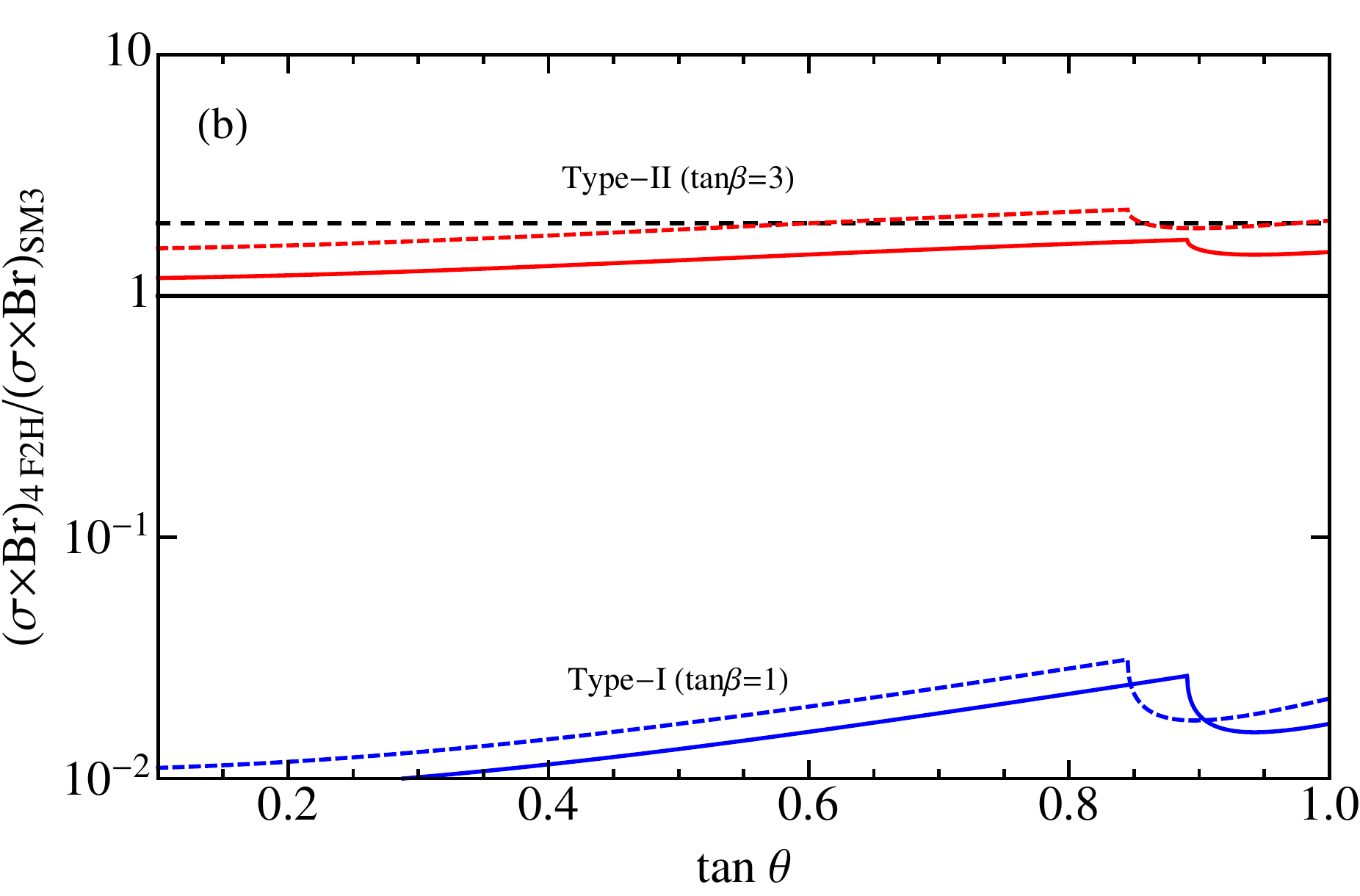}
\caption{Predicted $\gamma\gamma$ signals from the CP-odd Higgs boson $A^0$ at the LHC,
computed as the ratio
$\,(\sigma\times\text{Br})_{\rm 4F2H}^{}/(\sigma\times\text{Br})_{\rm SM3}^{}$.
Plot-(a) shows this ratio versus $\,\tanb\,$,\, for $\,\tant=1\,$.\,
Plot-(b) displays the same ratio as a function of $\,\tant$\,,\,
for $\,\tanb=3\,$ (red curves) and $\,\tanb=1\,$ (blue curves).\,
For both type-I and type-II, we show the predictions for two different $A^0$ masses: $\,M_{A}=126\GeV$\, (solid curve) and $\,M_{A}=140\GeV$ (dashed curve).
The horizontal black solid-line indicates that the prediction coincides with the SM3
(with the same Higgs mass), while the black dashed-line denotes the signal prediction
being twice of the SM3.
}
\label{fig:A0gaga-sBr}
\end{figure}

In Fig.\,\ref{fig:BRA-tanB-tanT}(a), we analyze the decay branching fractions
of $\,A^0\,$ versus the mixing angle $\,\tan\theta\,$ of the fourth-family neutrinos.
It shows that the invisible decay channel always dominates over all other channels
for the full allowed range  $\,0.1\leqq \tan\theta\leqq 1\,$.\,
The branching ratios for all other SM decay channels are maximized around
$\,\tan\theta\simeq 0.88\,$,\, at which the second invisible decay channel $\,A^0\to\mN_4^{}\bar\mN_4^{}$\, is kinematically forbidden.
In Fig.\,\ref{fig:BRA-tanB-tanT}(b),
we further analyze various decay branching fractions
versus $\,\tanb$\, for the 4F2HDM-II. The invisible decay branching ratio \,Br[$A^0\to\Nu\bar{\nu}_4^{}, \mN_4\bar\mN_4$]\, is maximized for small $\,\tanb\simeq 1$\,.\,
When $\,\tanb$\, gets larger, it gets reduced and no longer dominates over other channels;
this could potentially cause too large $\gamma\gamma$ signals at the LHC for the 4F2HDM-II.
Such a danger is absent for the 4F2HDM-I,
where all the partial decay widths are suppressed by $\,\cot^2\!\beta\,$
due to $\,\Gamma_{\rm 4F2H-I}^{}\propto\cot^2\beta$\,.\,
Besides, in the 4F2HDM-I the production cross section of $\,A^0$\,
gets suppressed for larger $\,\tanb$\,,\, as shown in Fig.\,\ref{fig:ggA}.

Now we are ready to evaluate the signal predictions of
$\,(\sigma\times\text{Br})\,$ for $\,gg\to A^0\to \ga\ga\,$
in the 4F2HDM (type-I and type-II), and then derive the ratio
$\,(\sigma\times\text{Br})_{\rm 4F2H}^{}/(\sigma\times\text{Br})_{\rm SM3}^{}$\,,\,
where the denominator is the corresponding signals
$\,gg\to h^0\to \ga\ga\,$ in the SM3 with the same input of Higgs mass as $\,A^0\,$.\,
We present our results in Fig.\,\ref{fig:A0gaga-sBr} for the LHC\,(7\,TeV).
It shows that for the 4F2HDM-I, the predictions are always significantly
below that of the SM3.
For the 4F2HDM-II with $\,M_{A}=126\GeV$,\,
we find that the $\,\ga\ga$\, signals can be moderately larger than that
of the SM3 in the parameter region, $\,2.4\lesssim\tanb\lesssim 4$\, with $\,\tant=1$\, [Fig.\,\ref{fig:A0gaga-sBr}(a)], or $\,0.1\lesssim\tant\lesssim 1$\, with $\,\tanb=3$\, [Fig.\,\ref{fig:A0gaga-sBr}(b)].  But, larger values of $\,\tanb\gtrsim 4$\, in Fig.\,\ref{fig:A0gaga-sBr}(a) would cause too much excess of $\,\ga\ga\,$ signals,
and are excluded by the current data. With the $4.9\,\ifb$ data set,
ATLAS collaboration observed $\,2.8\sigma$\, excess of $\,\ga\ga$\, events
at the invariant-mass $\,M_{\ga\ga}=126\,$GeV,\,
while the expected SM Higgs signal is $\,1.4\sigma$\, above the SM backgrounds,
which is about a factor-2 smaller than what ATLAS observed\,\cite{Atlas2011-12}.
The signal-reduction-rate due to various cuts and
detection efficiencies should be roughly the same
for both the SM Higgs boson and the CP-odd $\,A^0\,$ boson.
When the predicted ratio
$\,(\sigma\times\text{Br})_{\rm 4F2H}^{}/(\sigma\times\text{Br})_{\rm SM3}^{}\simeq 2$\,
in the 4F2HDM-II, the $\,A^0\to \ga\ga$\, signals can nicely
explain the $\,2.8\sigma$\, excess of ATLAS observation at $\,M_{\ga\ga}=126\,$GeV.\,
For instance, this is realized at $\,(\tanb,\,\tan\theta)\simeq (3.5,\,1)$\, in Fig.\,\ref{fig:A0gaga-sBr}(a).

As shown by Figs.\,\ref{fig:BRA-mA}-\ref{fig:BRA-tanB-tanT},
the invisible decays \,$A^0\to \Nuu,\mN_4\mN_4$\,
can dominate over all other channels in the relevant parameter regions.
The suppression on the fermionic decay branching ratios
(such as $\,A^0\to b\bar b,\,\tau\bar\tau$\,) appears moderate in comparison with
the SM3 case. Nevertheless, the Higgs searches
in the $\,b\bar b$\, and $\,\tau\bar\tau$\,
decay modes\,\cite{CMS2011-12} are made through the vector boson
associated production and the vector boson fusion processes, respectively,
which receive no new enhancement from the fourth-family fermions.
This is consistent with the present observations of ATLAS\,\cite{Atlas2011-12} and CMS\,\cite{CMS2011-12}, which found no excess from the
$\,b\bar b$\, and $\,\tau\bar\tau$\, final states.
Similar reasoning also holds for the detection of the CP-even Higgs boson $\,h^0$\, via the
$\,b\bar b$\, and $\,\tau\bar\tau$\, channels (cf.\ Sec.\,\ref{sec:3}).
Furthermore, the CMS detector showed no new signals in the $VV^*$ final states,
and ATLAS analysis only indicated a smaller excess in the $VV^*$ events.
It is very likely that the $VV^*$ channels contains only the SM backgrounds.
If so, this is again consistent with our analysis of the $\,A^0\,$ Higgs boson,
since the CP-odd $\,A^0\,$ has no tree-level gauge couplings with $VV$ and
thus the $\,A^0\to VV^*$\, decays are forbidden.

Note that in each plot of Fig.\,\ref{fig:A0gaga-sBr}, the two solid curves correspond to $\,M_A=126$\,GeV,\, and the two dashed curves to $\,M_A=140$\,GeV.\,
We do not show a curve for $M_A=116$\,GeV since it almost overlaps with that of
$\,M_A=126$\,GeV.\,  So the parameter space between the two adjacent curves in each set
(either red or blue) in Fig.\,\ref{fig:A0gaga-sBr} essentially
represent that of the mass-range $\,116\lesssim M_A^{}\lesssim 140\,$GeV.\,
From Fig.\,\ref{fig:A0gaga-sBr},
we see that should the present ATLAS excess at $\,M_{\ga\ga}=126\,$GeV
be disconfirmed by this summer with more LHC data,
our 4F2HDM-II can predict new Higgs signals in other
$\,M_{\ga\ga}$\, values around $116-140$\,GeV, either above or below the SM3 Higgs rates.
This will be further probed by the LHC Higgs searches.


\vspace*{4mm}
\section{\hspace*{-1mm}Signals of CP-Even ${\mathbf h}^{\mathbf 0}$
                       in 4F2HDM with Invisible Decays}
\vspace*{2mm}
\label{sec:3}

In this section, we turn to the analysis of the CP-even Higgs boson $\,h^0$\,
in the 4F2HDM, and study the impacts of the invisible decays $\,h^0\to\Nuu,\NNN\,$
on the LHC discovery. Due to the mixing between the two CP-even neutral states, we have the mixing angle $\a$ as a new input parameter. Unlike $A^0$,\,
the CP-even Higgs boson $\,h^0$\, also has additional decay channels of
$\,h^0\to WW^*,\,ZZ^*\,$ at tree-level. We will present a benchmark model for the 4F2HDM, and analyze the production and decays of $\,h^0\,$ at the LHC. We further compare our predictions to that of the SM Higgs coupled with
four families of fermions (SM4), by including the invisible decay channel
of $\,h^0\,$.\,

\begin{table}[h!]
\vspace*{4mm}
\begin{center}
\begin{tabular}{c||c|c}
\hline
\hline
   & ~4F2HDM-I~ & ~4F2HDM-II~     \\\hline
 $\xi_h^u$  & $\cos\alpha/\sin\beta$ & $\cos\alpha/\sin\beta$   \\
 $\xi_h^d$  & $\cos\alpha/\sin\beta$ & $-\sin\alpha/\cos\beta$~~   \\
 $\xi_h^{\nu}$  & $\cos\alpha/\sin\beta$ & $\cos\alpha/\sin\beta$   \\
 $\xi_h^{\ell}$  & $\cos\alpha/\sin\beta$ & $-\sin\alpha/\cos\beta$~~  \\
  \hline
\hline
\end{tabular}
\caption{Yukawa couplings of the SM fermions to the lighter CP-even Higgs boson $h^0$
for the 4F2HDM-I and 4F2HDM-II.}
\label{tab:2HDSM4-hYukawa}
\end{center}
\end{table}

In the 4F2HDM, the analysis of production and decays of $\,h^0$\,
are more complicated than $\,A^0\,$,\, due to the additional decay channels
in the $\,WW$\, and $\,ZZ$\, final states, as well as the mixing parameter $\a$
associated with two CP-even states $(h^0, H^0)$. The Yukawa interactions of $\,h^0$\,
can be generally expressed as follows,
\beqn
\label{eq:Y-hff}
\mL_{\rm Yukawa}^{} &\,=\,&- \sum_f\frac{\,m_f^{}\,}{v}\xi_h^f\,\bar{f}f\, h^0 \,,
\eeqn
where the Yukawa couplings $\xi_h^f$ for the 4F2HDM-I and 4F2HDM-II are summarized in Table\,\ref{tab:2HDSM4-hYukawa}.
The $\,h^0$\, production via the gluon-fusion process receives
new contributions from the fourth-family quarks $(t_4^{}, b_4^{})$,
which are enhanced by the Yukawa couplings of $(t_4^{}, b_4^{})$
relative to that of the SM top quark ($t$). We compute the ratio of
production cross sections between the 4F2HDM and SM3
with the same mass of $\,h^0\,$,
\beqn
\label{eq:ggh-enhance}
\frac{~\sigma[gg\to h^0]_{\rm 4F2H}^{}~}{\sigma[gg\to h^0]_{\rm SM3}^{}}
~=~
\frac{\,\dis\Big|\sum_{Q=t,t_4^{}, b_4^{}} \xi_h^Q\,\mI_S^{}(\tau_Q^{})\Big|^2\,}
     {\left|\mI_S^{}(\tau_t^{})\right|^2} \,,
\eeqn
which is found to be generally larger than unity.
In Fig.\,\ref{fig:ggh}, we present the enhancement factors (\ref{eq:ggh-enhance})
for 4F2HDM-I (blue curve) and 4F2HDM-II (red curve) with the sample input $\,(\tanb,\,\tan\alpha)=(1,\,-3)$.\,
The enhancement (\ref{eq:ggh-enhance}) is moderate
since $\,|\xi_h^Q|$\, in (\ref{eq:ggh-enhance}) can be smaller than one,
as compared to the SM4 with $\,|\xi_h^Q|=1$.\,
For the SM4, we see from  the purple curve in Fig.\,\ref{fig:ggh},
$\,\sigma[gg\to h^0]_{\rm SM4}^{}/\sigma[gg\to h^0]_{\rm SM3}^{}\approx 9$\,
holds in the limit of light Higgs mass $\,\Mhh\ll (2M_Q^{})^2$\,
with the loop-contributions from the heavy quarks.

\begin{figure}
\centering
\includegraphics[width=10cm,height=8cm]{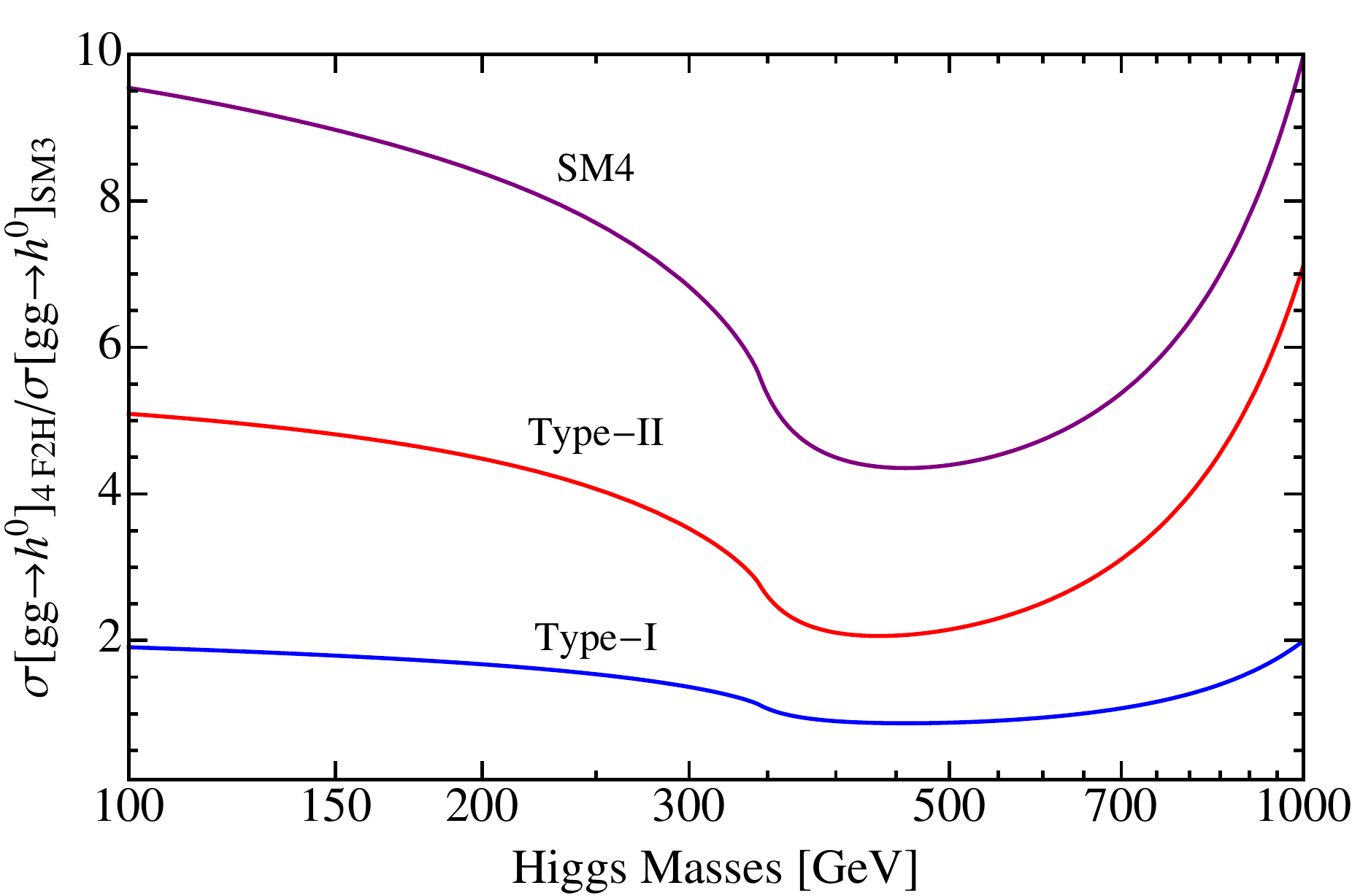}
\caption{Predicted ratio $\,\sigma[gg\to h^0]_{\rm 4F2H}^{}/\sigma[gg\to h^0]_{\rm SM3}^{}$\,
for the 4F2HDM-I (blue curve) and 4F2HDM-II (red curve) with the sample input
$\,(\tanb,\,\tan\alpha)=(1,\,-3)$.\, As a comparison, the ratio $\,\sigma[gg\to h^0]_{\rm SM4}^{}/\sigma[gg\to h^0]_{\rm SM3}^{}$\, is shown for the SM4 (purple curve).}
\label{fig:ggh}
\end{figure}

\begin{figure}
\centering
\includegraphics[width=12cm,height=9.5cm]{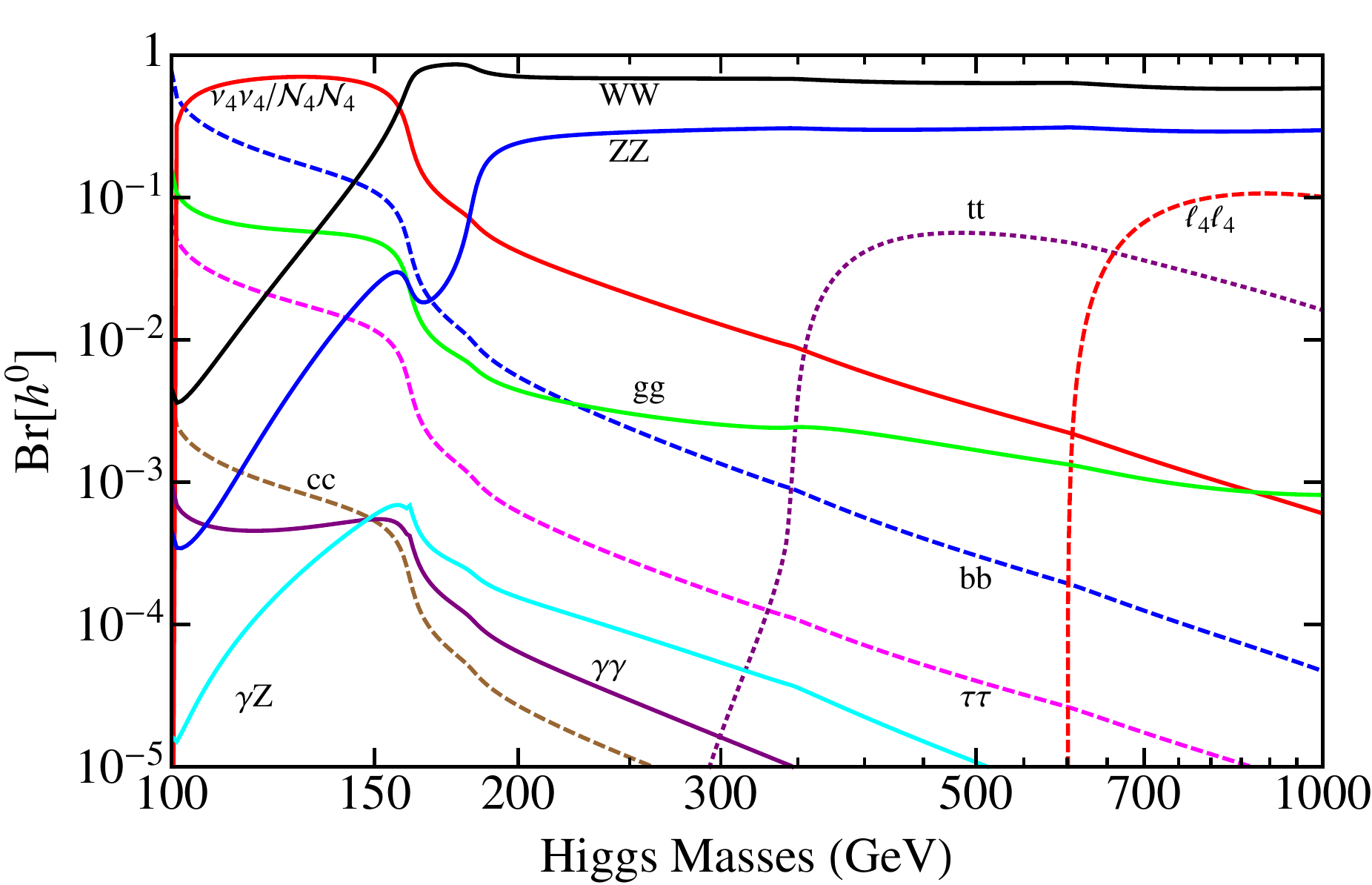}
\caption{Decay branching ratios of CP-even Higgs boson $\,h^0\,$ in the 4F2HDM-II
for the mass-range $\,M_h=100-1000$\,GeV\,
and sample input \,$(\tanb,\,\tan\alpha)=(1,\,-3)$\,.}
\label{fig:Br-h0-II}
\vspace*{10mm}
\end{figure}

\begin{figure}[h]
\centering
\includegraphics[width=10.5cm,height=8.5cm]{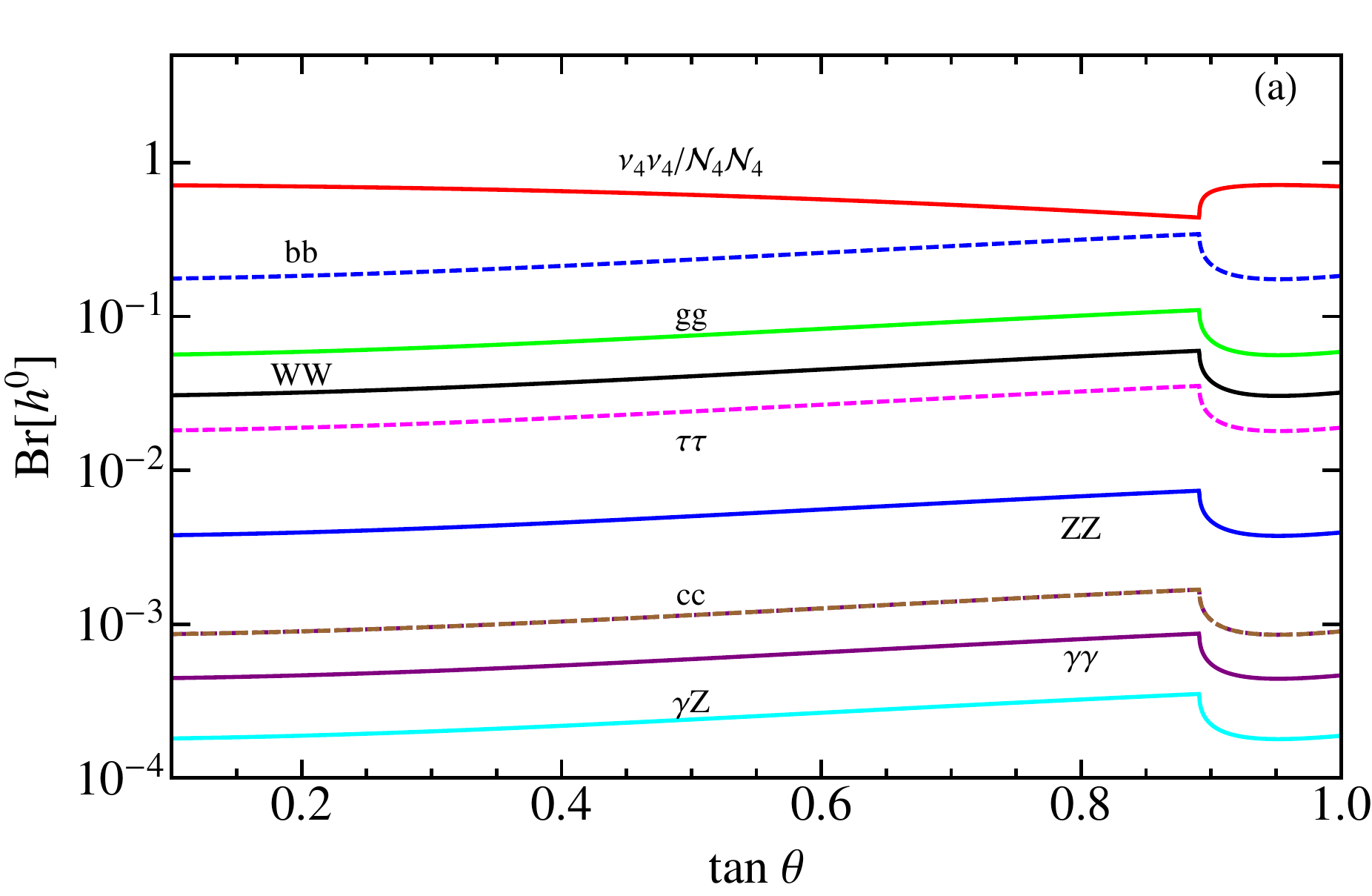}
\\[2mm]
\includegraphics[width=10.5cm,height=8.5cm]{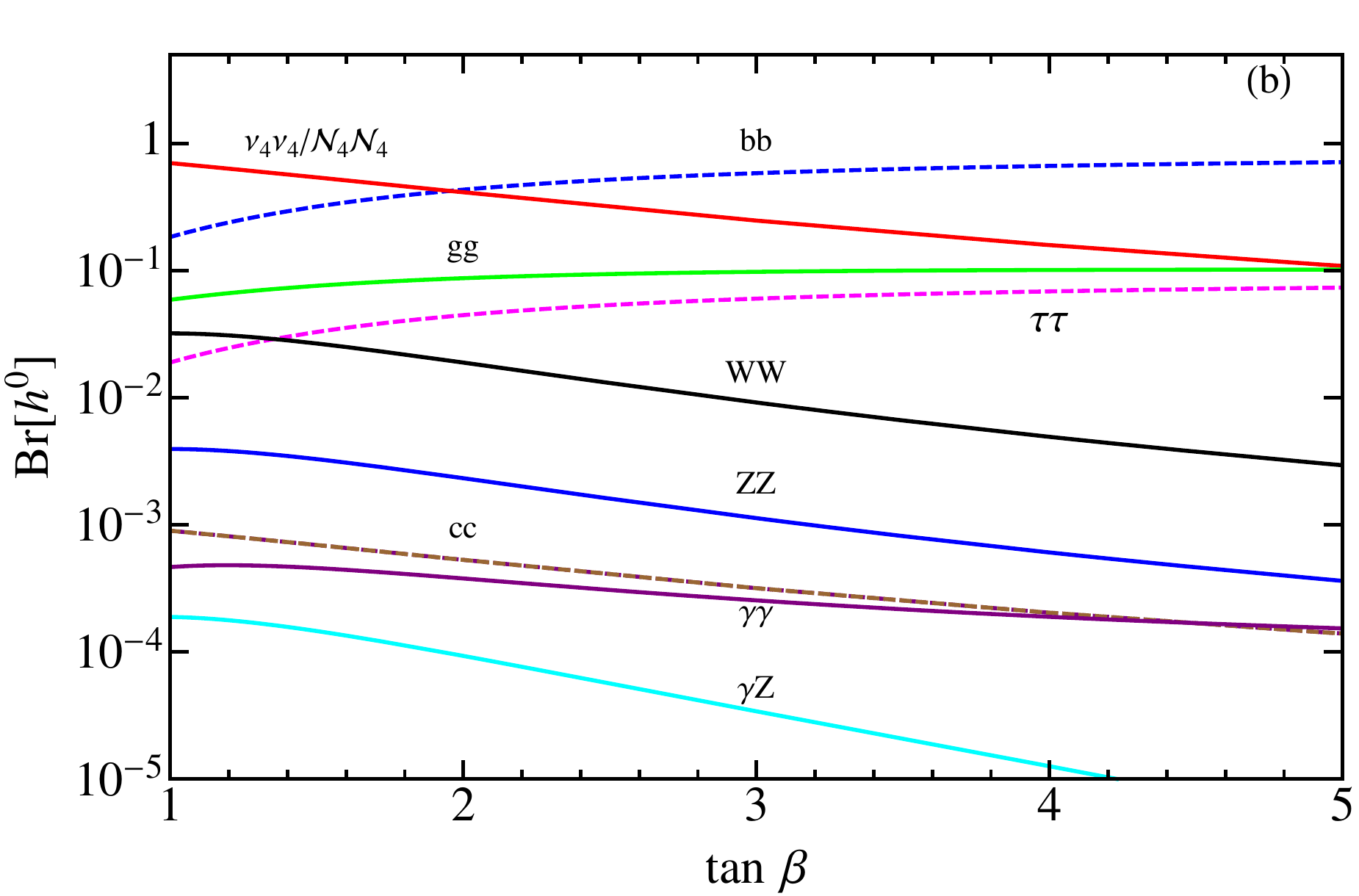}
\caption{Decay braching ratios of CP-even Higgs boson $h^0$ with mass $\Mh=126$\,GeV. Plot-(a): Br$[h^0]$ versus $\tan\theta$ with $\tanb =1$.\, Plot-(b): Br$[h^0]$ versus $\tanb$, with $\tan\theta=1$. In both plots $\,\tan\alpha=-3.0$\, is fixed.}
\label{fig:h0-Br-tanB-tanT}
\end{figure}

We analyze the impact of fourth-family fermions on the Higgs boson $\,h^0$\, decays.
For relatively light fourth-family neutrinos and leptons,
new decay channels of $\,h^0\to \Nuu, \NNN, \ell_4^{}\bar\ell_4^{}$\,
can be open, in addition to the conventional SM decay modes whose partial widths
have rescaling factors $\,|\xi_h^f|^2$\, and $\,\sin^2(\beta-\alpha)$\,
for the fermionic and $VV^*$ final states, respectively.
Therefore, we shall rewrite the loop-induced decay rates in terms of
the modified couplings for both bosonic and fermionic contributions,
\beqa
\Gamma(h^0\to gg)_{\rm 4F2H}^{} &=&
\frac{\alpha_s^2 M_h^3}{8\pi^3 v^2}
\Big|\sum_{Q=t,t_4^{}, b_4^{}} \!\!\! \xi_h^Q\,\mI_S^{}(\tau_{Q}^{})\Big|^2 ,
\label{hgg_rate}
\eeqa
\beqa
\Gamma(h^0\to\gamma\gamma)_{\rm 4F2H}^{} &=&
\frac{\alpha^2 M_h^3}{16\pi^3 v^2}\Big|\sum_{f=t,t_4^{}, b_4^{},\ell_4^{}} \!\!\!\!\!\!
N_{c}^f e_f^2\xi_h^f\mI_S^{}(\tau_f^{})
+ \frac{1}{2}\sin(\beta\!-\!\alpha)\,\mI_W^{}(\tau_W^{})\Big|^2 ,
\hspace*{17mm}
\label{hgaga_rate}
\\[2mm]
\Gamma(h^0\to\gamma Z^0)_{\rm 4F2H}^{} &=&
\frac{\alpha M_h^3 m_W^2 }{128\pi^4 v^4}\left(\!1-\frac{\,m_Z^2}{\Mhh}\right)^{\!3}
\,\Big|\sum_{f=t,t_4^{},b_4^{},\ell_4^{}} \!\!\!\!\!\!\xi_h^f N_{c}^f
  \frac{e_f^{}c_f^{}}{c_W^{}}\mA_f^H(\tau_f^{}, \lambda_f^{})
\nn\\
&& +\sin(\beta\!-\!\alpha)\,\mA_W^H(\tau_W^{},\lambda_W^{})\Big|^2 ,
\label{hzga_rate}
\eeqa
with the form factors,
\beqs
\beqa
\mI_W^{}(\tau) &=&  -\frac{1}{\tau^2}\Big[2\tau^2+3\tau+3(2\tau-1)f(\tau) \Big] ,
\\[2mm]
\mA_f^H(\tau, \lambda) &=&  I_1^{}(\tau,\lambda)-I_2^{}(\tau,\lambda) \,,
\\[2mm]
\mA_W^H(\tau,\lambda) &=&  c_W^{}
\left\{ 4\(3-\frac{s_W^2}{c_W^2}\)I_2^{}(\tau,\lambda)
+\left[(1+2\tau)\frac{s_W^2}{c_W^2}-(5+2\tau)\right]I_1^{}(\tau,\lambda)\right\} ,
\hspace*{14mm}
\\[2mm]
I_1^{}(\tau,\lambda ) &=&
\frac{1}{\,2(\lambda-\tau)\,}
+\frac{\,f(\tau)-f(\lambda)\,}{\,2(\lambda-\tau)^2\,}
+\frac{\,\lambda \left[ g(\tau)-g(\lambda)\right]\,}{(\tau-\lambda)^2} \,,
\\[2mm]
I_2^{}(\tau,\lambda) &=&  \frac{\,f(\tau) -f(\lambda)\,}{2(\tau-\lambda)} \,,
\eeqa
\beqa
g(\tau) &=& \left\{
\ba{ll}
\sqrt{\tau^{-1}-1}\arcsin\sqrt{\tau} \,, ~~&~~ \tau\leqq 1 \,,
\\[2.5mm]
\dis\frac{\sqrt{1-\tau^{-1}}}{2}
\left[\ln\frac{1\!+\!\sqrt{1\!-\!\tau^{-1}}}{1\!-\!\sqrt{1\!-\!\tau^{-1}}}-i\pi\right] ,
~~&~~ \tau>1 \,.
\ea \right.
\eeqn
\eeqs
The charged Higgs loops may also contribute to the (\ref{hgaga_rate}) and (\ref{hzga_rate}).
For our illustration, we consider the large $M_{\pm}$ limit where the $H^{\pm}$
contributions are negligible.
Such large $M_{\pm}$ limit is also fairly reasonable from the flavor physics constraints,
including the leptonic decay of mesons $M\to\ell\nu$, loop-induced $b\to s\gamma$ transitions,
and the mass difference $\Delta M_B$ as measured
in the $B^0-\bar B^0$ mixing \cite{Mahmoudi:2009zx, Deschamps:2009rh}.
For lighter $H^{\pm}$,\, the inclusion of charged Higgs loop
will not affect our physical conclusion. In Fig.\,\ref{fig:Br-h0-II},
we presented a sample of decay branching fractions for $\,h^0$\,
as a function of its mass $\Mh$ in the 4F2HDM-II,
by including the new invisible decay modes.
It clearly shows that the invisible decays $\,h^0\to \Nuu,\mN_4^{}\mN_4^{}$\,
can suppress the other decay channels in the light mass region of $\,M_h\lesssim 2M_W^{}$.\,
For $\,M_h\gtrsim 160\GeV$,\, they no longer dominate because of
\,$\Gamma(h^0\to\Nuu, \mN_4\mN_4)/\Gamma(h^0\to VV) <1$\,.\,
From Fig.\,\ref{fig:ggh}-\ref{fig:Br-h0-II},  we see that for a light $\,h^0$\,
with mass  $\,116\GeV < M_h < 2m_W^{}\,$,\, its production and decays in the 4F2HDM
are very different from that of the SM Higgs boson due to the fourth-family quark
contributions and the new channels of invisible decays.

In Fig.\,\ref{fig:h0-Br-tanB-tanT}(a), we present the decay branching fractions of
$\,h^0$\, as a function of $\,\tan\theta$\, with \,$\tanb =1$\,,\,
while in Fig.\,\ref{fig:h0-Br-tanB-tanT}(b) we display \,Br$[h^0]$\,
as a function of $\,\tanb$\, with \,$\tan\theta=1\,$.\,
We find that \,Br$[h^0]$\, is sensitive to $\,\tanb\,$,\,
and for $\,\tanb\lesssim 2\,$ the invisible decays
$\,h^0\to\Nuu,\NNN\,$ dominate over all other SM channels
in the full range of $\,\tan\theta$\,.

\begin{figure}[t]
\centering
\includegraphics[width=10.5cm,height=9cm]{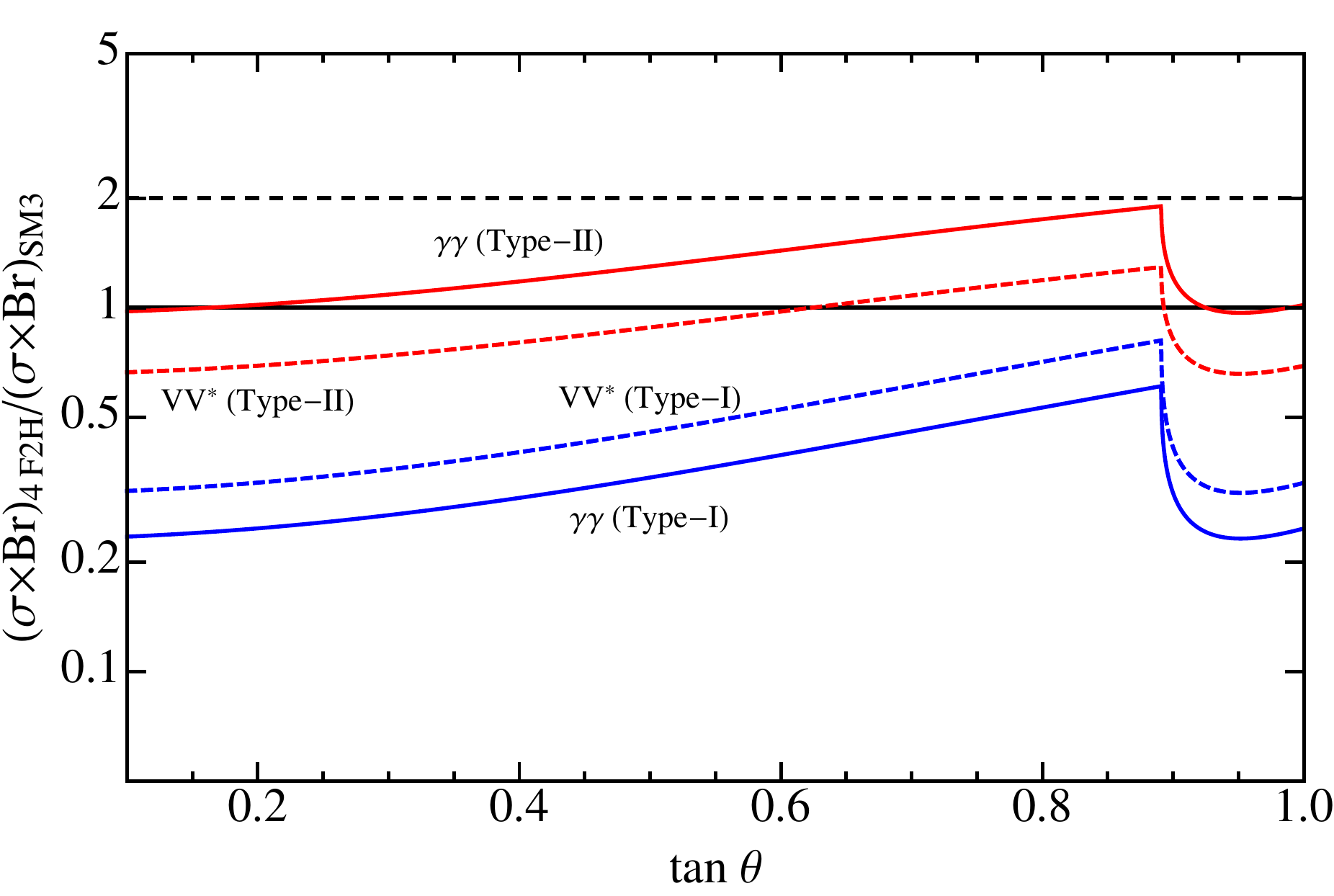}
\vspace*{-2mm}
\caption{Predicted signal ratios of $\,\(\sigma\!\times\!\text{Br}\)_{\rm 4F2H}^{}/\(\sigma\!\times\!\text{Br}\)_{\rm SM3}^{}$
in the processes $\,gg\to h^0\to\gamma\gamma$\,
and $\,gg\to h^0\to VV^*$\, as functions of $\,\tan\theta$\,,\,
with Higgs mass $\,M_h=126\GeV$ for both 4F2HDM-I and 4F2HDM-II.
We have take the sample inputs \,$(\tanb,\,\tan\alpha)=(1,\,-3)$.\,}
\label{fig:h0-tanT-CSBr}
\end{figure}
\begin{figure}[t]
\centering
\includegraphics[width=10.3cm,height=8.3cm]{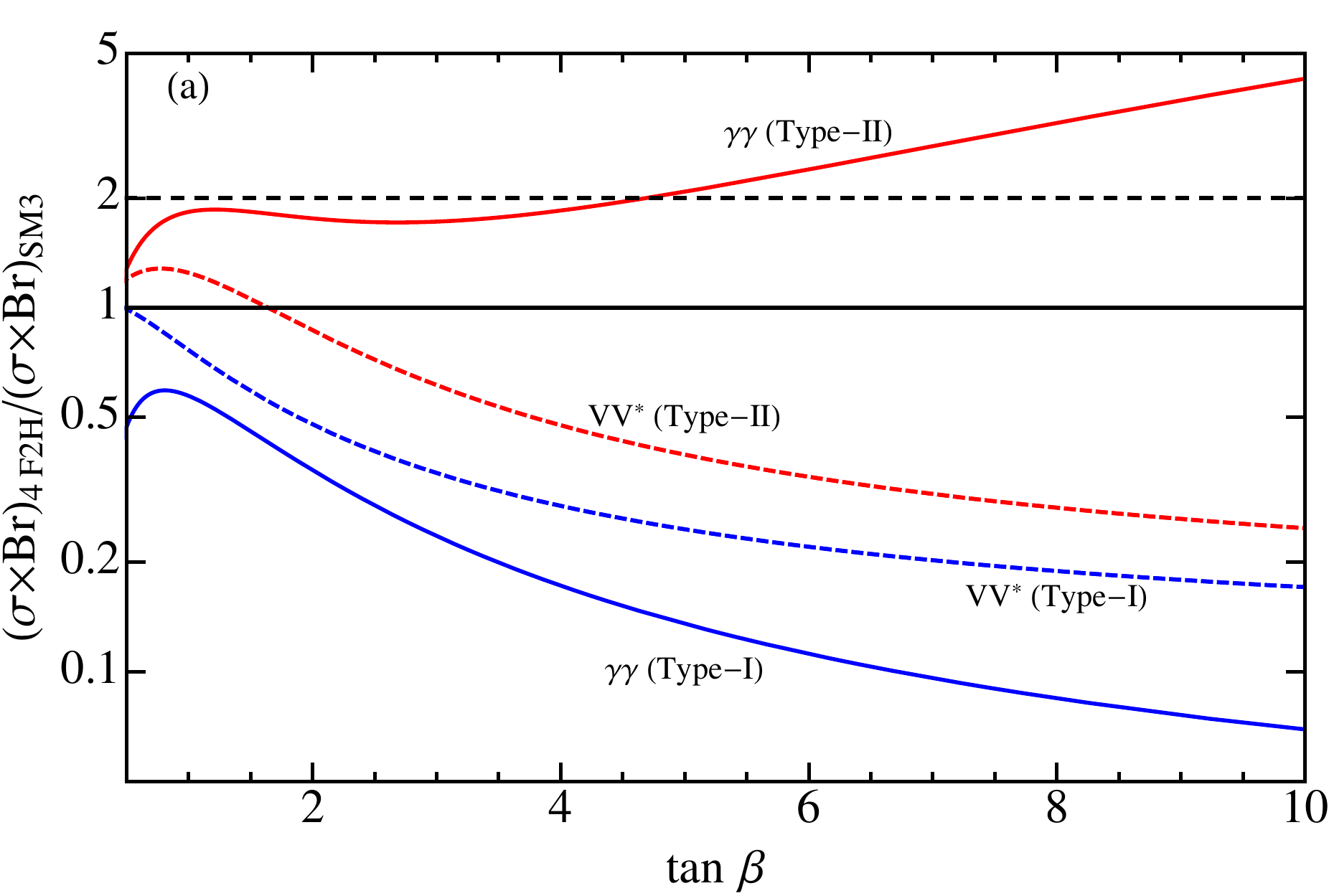}
\\[2mm]
\includegraphics[width=10.3cm,height=8.3cm]{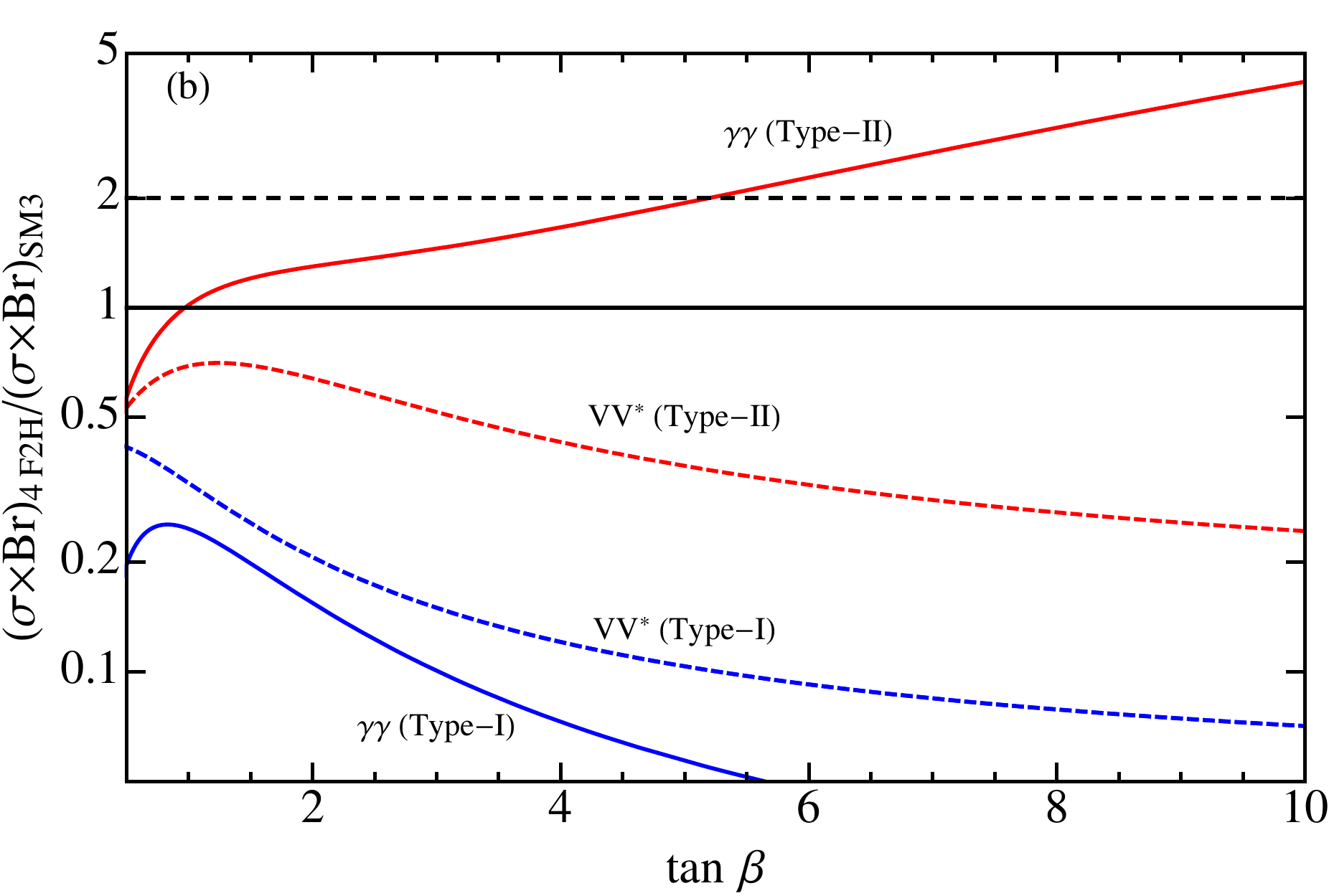}
\caption{Predicted signal ratios of $\,\(\sigma\!\times\!\text{Br}\)_{\rm 4F2H}^{}/\(\sigma\!\times\!\text{Br}\)_{\rm SM3}^{}$ for $\,gg\to h^0\to\gamma\gamma$\,
and $\,gg\to h^0\to VV^*$\, as functions of $\,\tanb$,\, with Higgs mass $\,M_h=126\GeV$ for both 4F2HDM-I and 4F2HDM-II. The neutrino mixing angle is taken to be $\,\tan\theta=0.8$\, in plot-(a) and $\,\tan\theta=1.0$\, in plot-(b), while the Higgs mixing angle $\,\tan\alpha =-3.0$\, in both plots. The horizontal dashed-line in each plot corresponds to the ratio, $\,\(\sigma\!\times\!\text{Br}\)_{\rm 4F2H}^{}/\(\sigma\!\times\!\text{Br}\)_{\rm SM3}^{}=2$\,.}
\label{fig:h0-tanb-CSBr}
\end{figure}

\begin{figure}
\centering
\includegraphics[width=10.3cm,height=8.3cm]{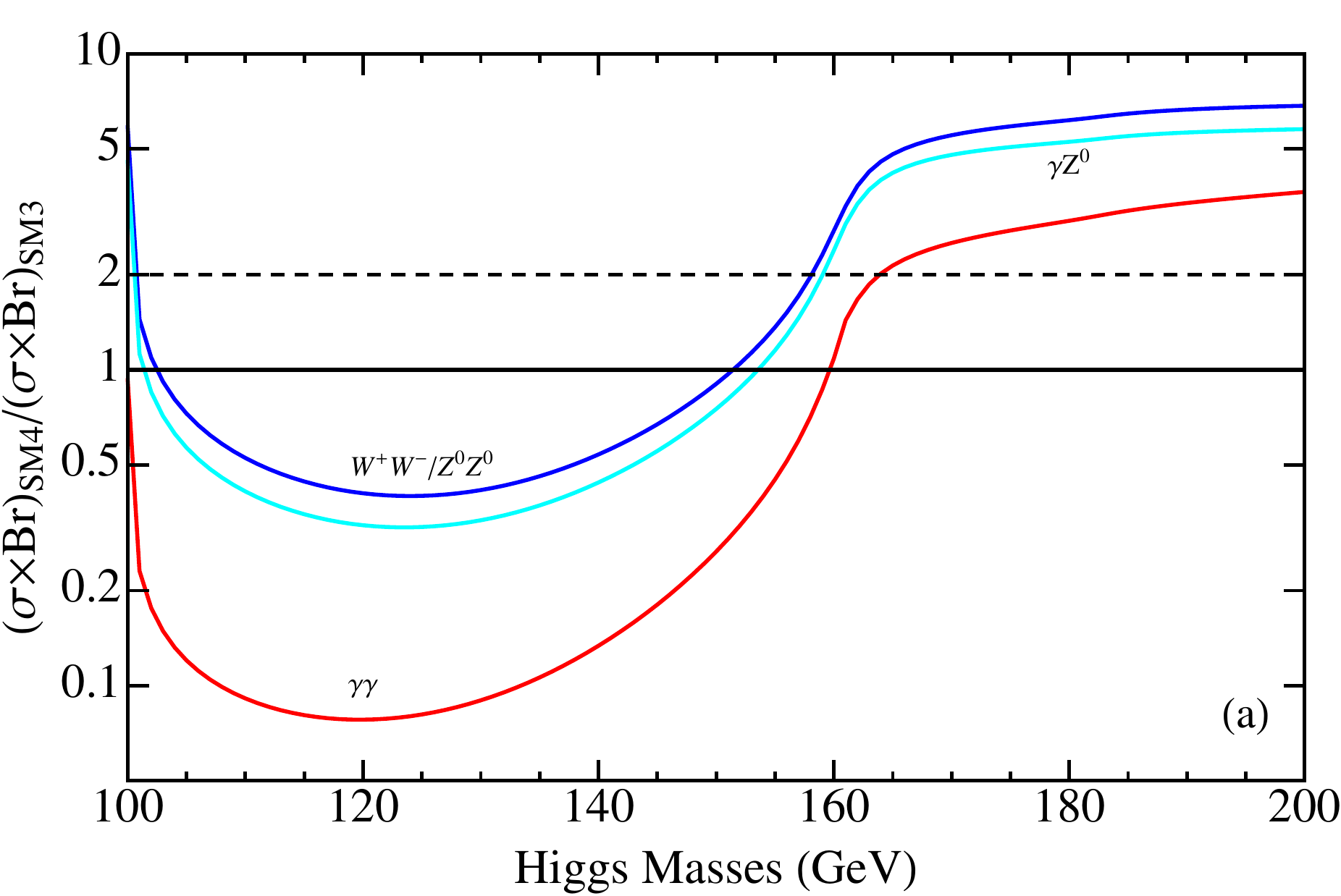}
\\[2mm]
\includegraphics[width=10.3cm,height=8.3cm]{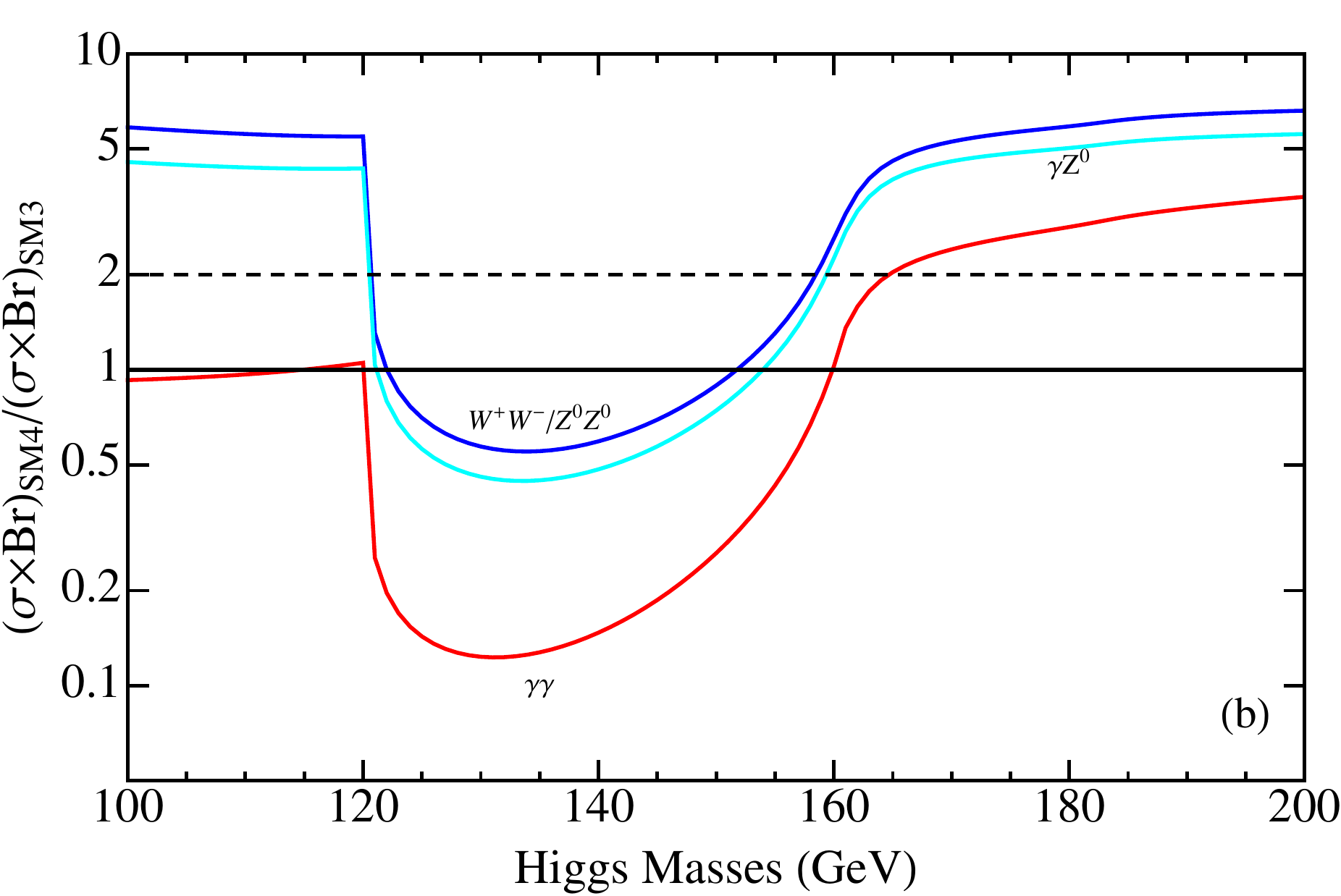}
\caption{Predicted signal ratios of $\,\(\sigma\!\times\!\text{Br}\)_{\rm SM4}^{}/\(\sigma\!\times\!\text{Br}\)_{\rm SM3}^{}$ in the processes
$\,gg\to h^0\to\gamma\gamma$\, and $\,gg\to h^0\to VV^*$\,
as functions of Higgs mass $\,M_h$\, for the SM4. The fourth-family neutrino mass $\,M_{\nu_4^{}}^{}=50$\,GeV in plot-(a) and $\,M_{\nu_4^{}}^{}=60$\,GeV in plot-(b)
with $\,\tan\theta=1\,$ are taken.}
\label{fig:h0-SM4-CSBr}
\end{figure}

In Fig.\,\ref{fig:h0-tanT-CSBr}-\ref{fig:h0-tanb-CSBr},
we present the predicted $\sigma\!\times\!{\rm Br}$
for the processes $\,gg\to h^0\to\gamma\gamma$\, and $\,gg\to h^0\to VV^*$\,
in both 4F2HDM-I and 4F2HDM-II.\, In particular, we show the ratio $\,\(\sigma\!\times\!\text{Br}\)_{\rm 4F2H}^{}/\(\sigma\!\times\!\text{Br}\)_{\rm SM3}^{}$\,,\,
for the comparison to that of the SM Higgs boson, with the same mass $\,M_h^{}=126\,$GeV.\,
In Fig.\,\ref{fig:h0-tanT-CSBr}, we plot the ratios
$\,\(\sigma\!\times\!\text{Br}\)_{\rm 4F2H}^{}/\(\sigma\!\times\!\text{Br}\)_{\rm SM3}^{}$\,
for the 4F2HDM-I and 4F2HDM-II as functions of the neutrino mixing parameter $\,\tan\theta\,$,\,
with a sample input of \,$(\tanb,\,\tan\alpha)=(1,\,-3)$.\,
The predictions of the 4F2HDM-I are generally suppressed in comparison with the SM3,
thus they cannot be observed from the current LHC data.
To detect $\,h^0\,$ in the 4F2HDM-I thus requires
higher integrated luminosities at the LHC.
For predictions of the 4F2HDM-II, Fig.\,\ref{fig:h0-tanT-CSBr}
shows interesting excess of signals above that of the SM Higgs boson
for both $\,\ga\ga\,$ and $\,VV^*$\, channels
in the parameter range $\,0.63\leqq \tan\theta \leqq 0.9\,$
(with $\tanb=1$), where the $\,\ga\ga$\, signals are
significantly higher than the $VV^*$ signals.\,
We note that such 4F2HDM-II model with $\,\tanb\sim 1\,$ is quite generic
for the dynamical fourth-family models  \cite{Burdman:2011fw}.
Combined with the invisible decay channels, this can nicely explain
why ATLAS experiment\,\cite{Atlas2011-12} has detected sizable excess of events
in the $\,\ga\ga$\, mode but not the $ZZ^*$ and $WW^*$ final states.
Fig.\,\ref{fig:h0-tanb-CSBr}(a)-(b) depict these signal ratios
as functions of $\,\tanb\,$ for the neutrino mixing parameter
$\,\tan\theta=0.8\,$ and $\,\tan\theta=1.0\,$, respectively.
 We find that the 4F2HDM-II always predicts larger signals than the SM3
 in the $\,\ga\ga$\, channel for $\,\tanb \gtrsim 1\,$,\,
 while the $\,VV^*$\,  signals are mainly suppressed
 except for $\,\tanb\lesssim 2\,$ (with $\tan\theta=0.8$).\,

For comparison with our above 4F2HDM studies, we also analyze the $\,h^0\,$ signals in the $\,\gamma\gamma$\, and $\,VV^*$\, channels
from the one-Higgs-doublet SM including fourth-family (SM4),
with relatively light $\Nu/\NN$\,.\,
Ref.\,\cite{Guo:2011ab} showed $\ga\ga/VV^*$ signals from $h^0$ in the SM4
without including invisible decays, while the effect of invisible decays
for the SM4 Higgs was discussed in \cite{Belotsky:2002ym} for the LEP searches and in \cite{Carpenter:2011wb, Keung-Cetin:2011} for the LHC searches.
In Fig.\,\ref{fig:h0-SM4-CSBr}, we present the predicted signal ratios of $\,\(\sigma\!\times\!\text{Br}\)_{\rm SM4}^{}/\(\sigma\!\times\!\text{Br}\)_{\rm SM3}^{}$\,
in the processes $\,gg\to h^0\to\gamma\gamma$\, and $\,gg\to h^0\to VV^*$\,
as functions of Higgs mass $\,M_h$\, for the SM4.
For comparison, we assign the fourth-family neutrino mass, $\,M_{\nu_4^{}}^{}=50$\,GeV
in Fig.\,\ref{fig:h0-SM4-CSBr}(a) and $\,M_{\nu_4^{}}^{}=60$\,GeV
in Fig.\,\ref{fig:h0-SM4-CSBr}(b), respectively.
From these plots, we see that the $\,\gamma\gamma$\, signals are always
much more suppressed than the $\,VV^*$\, signals.
For the $\,\ga\ga$\, final states, besides the overall suppression from invisible decays $\,h^0\to\Nuu/\mN_4\mN_4$,\, the fourth-family fermions further suppress the decay width of $\,h^0\to\gamma\gamma$\, due to the enhanced fermion-loop contributions that cancel against the $W$-loop in (\ref{hgaga_rate}).
Hence, for the Higgs detection in the SM4,
the LHC should observe significantly larger $VV^*$ signals than
the $\gamma\gamma$ signals; this is just opposite to the most recent
ATLAS and CMS observations\,\cite{Atlas2011-12}\cite{CMS2011-12}.
Furthermore, Fig.\,\ref{fig:h0-SM4-CSBr}(a) shows that the SM4 predictions
for the light mass-range $\,105\lesssim M_h\lesssim 150\,$GeV are generally lower than
that of the conventional SM3 through all three decay channels of
$\,\ga\ga$,\, $\,\ga Z$\, and \,$VV^*$\,.\,
Hence, higher integrated luminosities at the LHC are required for its detection.
For Fig.\,\ref{fig:h0-SM4-CSBr}(b) with a larger fourth-neutrino mass
$\,M_{\nu_4^{}}^{}=60$\,GeV, this window shifts to
$\,120\lesssim M_h\lesssim 150\!-\!160\,$GeV.
Fig.\,\ref{fig:h0-SM4-CSBr}(b) shows that
the SM Higgs boson with $\,M_h < 120\,$GeV or $\,M_h > 160\,$GeV
is clearly excluded by the current LHC data due to excessive signals
in the \,$VV^*$\, final states. If the recent event excesses
around mass-values of $124-126$\,GeV at the LHC\,(7\,TeV)\,\cite{Atlas2011-12,CMS2011-12}
are actually due to statistical fluctuations or other systematical errors, then the low
Higgs-mass-ranges of the SM4, namely $\,105\lesssim M_h\lesssim 150\,$GeV in plot-(a) and $\,120\lesssim M_h\lesssim 150\,$GeV in plot-(b), are still viable and will be
further probed at the LHC with higher luminosities.

\vspace*{4mm}
\section{\hspace*{-1mm}Conclusions}
\vspace*{2mm}
\label{sec:4}

The on-going LHC Higgs searches for the light mass window ($116-130$\,GeV) are
crucial for testing the Higgs mechanism and probing new physics beyond the SM.
In this work, we studied the new signatures of a light CP-odd Higgs $\,A^0$\,
or CP-even Higgs $\,h^0$\, in the $\,\gamma\gamma$\, and $\,VV^*$\, channels ($V=W,Z$)
at the LHC\,(7\,TeV), as predicted by the two-Higgs-doublet-model
with the fourth-family fermions (4F2HDM).
{\it By including the invisible decays of Higgs boson $\,A^0$\, or $\,h^0$\,
into fourth-family neutrinos $\,\Nuu/\NNN\,$,}\,
we demonstrated that the predicted $\,\gamma\gamma$\, and \,$VV^*$\, signals
can explain the recently observed excesses of events
in ATLAS\,\cite{Atlas2011-12} and CMS\,\cite{CMS2011-12} detectors.
Due to the absence of cubic gauge-couplings \,$A^0$-$V$-$V$\,,\,
the decay channel $\,A^0\to\ga\ga$\, becomes unique for discovering $\,A^0$\,
in the light mass-range \,$116-130$\,GeV.\,
Although the fourth-family quark-loops significantly enhance
the production cross section of the gluon-fusion process
$\,gg\to A^0\,$ relative to that of $\,gg\to h^0\,$
in the conventional three-family SM (SM3) as in Fig.\,\ref{fig:ggA},
the invisible decay modes $\,A^0\to \Nuu,\NNN\,$ can properly suppress the
$\,A^0\to\ga\ga$\, branching fraction (Figs.\,\ref{fig:BRA-mA}-\ref{fig:BRA-tanB-tanT})
and make the $\,\ga\ga$\, signals mildly exceed that of the SM3 (Fig.\,\ref{fig:A0gaga-sBr}).
Hence, we found that for our 4F2HDM-II (with generic type-II Higgs sector),
a light $A^0$ with mass $124-126$\,GeV can nicely explain the excess of $\,\ga\ga$\,
signals at the LHC\,\cite{Atlas2011-12,CMS2011-12}.
At the same time, the $\,A^0$\, Higgs boson gives no signal for $\,VV^*$\, channels.
Note that the latest ATLAS search showed lower excesses in $\,VV^*$\, modes and
the CMS analysis found no excess in the same channel.
If a light Higgs boson indeed exists, more LHC data in 2012
will pin down the possible signals in both $\,\gamma\gamma$\, and $\,VV^*$\, channels,
and thus can further discriminate the CP-odd scalar $A^0$ from the CP-even scalar $h^0$.\,
In contrast, the 4F2HDM-I prediction of $\,A^0$\, signals in $\,\ga\ga\,$ mode
is always suppressed relative to that of the $\,h^0$\, in the SM3 throughout the parameter space.
Hence, for the 4F2HDM-I, detecting $\,A^0$\, in the $\,\ga\ga$\, channel will require
higher LHC luminosities than that of the SM3 or the 4F2HDM-II.

We further study a light CP-even Higgs boson $\,h^0\,$ in the 4F2HDM
(type-I and type-II) and analyze the LHC signals
in the presence of invisible decays $\,h^0\to \Nuu,\NNN\,$.\,
We demonstrated that the branching fraction of this invisible decay channel
becomes dominant in the light $\,h^0\,$ mass-range $116-150$\,GeV,\,
and the $\,\ga\ga$\, and $\,VV^*$\, modes are suppressed accordingly
in comparison with the SM3 (Fig.\,\ref{fig:Br-h0-II}-\ref{fig:h0-Br-tanB-tanT}).
Since fourth-family quarks always enhance the production cross section
of the gluon-fusion $\,gg\to h^0\,$ (Fig.\,\ref{fig:ggh}), we found that
for interesting parameter regions of the 4F2HDM-II, the final signals
(including decay branching fractions) can mildly exceed the SM Higgs signals
in the $\ga\ga$ channel, as well as the $VV^*$ channel with less enhancement
(cf.\ the parameter space $\,0.63 <\tan\theta < 0.9\,$ in Fig.\,\ref{fig:h0-tanT-CSBr}).
This can nicely explain the excesses of events for invariant-mass
$124-126$\,GeV as observed by ATLAS\,\cite{Atlas2011-12} and CMS\,\cite{CMS2011-12}.
We also found parameter regions with enhanced $\,\ga\ga\,$ signals, but suppressed
$\,VV^*\,$ events (Figs.\,\ref{fig:h0-tanT-CSBr}-\ref{fig:h0-tanb-CSBr}).
If the forthcoming LHC runs in 2012 only confirm the excess of $\,\ga\ga\,$ signals
but not the $\,VV^*$\, events, our model does provide good predictions for this,
as shown in Figs.\,\ref{fig:h0-tanT-CSBr}-\ref{fig:h0-tanb-CSBr}.
On the other hand, the 4F2HDM-I always predicts larger
$\,VV^*$\, signals than $\,\ga\ga$\,,\,
but they are both significantly lower than the SM3 and thus harder to detect
(Figs.\,\ref{fig:h0-tanT-CSBr}-\ref{fig:h0-tanb-CSBr}).
Should the present event excesses of ATLAS and CMS not be confirmed
with more data in 2012, the 4F2HDM-I will serve as a proper candidate
for this light Higgs mass-window. Thus, the on-going LHC Higgs searches
will probe the distinct predictions of the 4F2HDM.\footnote{
In passing, we also note that a recent study\,\cite{Englert:2011aa}
generally analyzed a hidden Higgs scenario with three-family fermions
for the LHC test, where the visible Higgs mixes with a hidden Higgs and
can have induced invisible decays.}

Finally, as a comparison with our 4F2HDM, we also analyzed light Higgs signals
for the one-Higgs-doublet SM with fourth-family fermions (SM4),
in the presence of invisible decay mode $\,h^0\to \Nuu,\NNN\,$.\,
We showed that due to the overall suppression on the decay branching fractions of
$\,\ga\ga$\, and $\,VV^*$\, final states, their predicted signal rates
are mostly below that of the SM3, as depicted in Fig.\,\ref{fig:h0-SM4-CSBr}(a)-(b)
for two sample inputs of the fourth-neutrino mass.
In addition, we found that the $\,VV^*\,$ rates are
always significantly higher than the $\,\ga\ga$\, signals.
These are very distinct from our 4F2HDM predictions
in Figs.\,\ref{fig:h0-tanT-CSBr}-\ref{fig:h0-tanb-CSBr},
and will be further probed by the forthcoming LHC runs in 2012.

\vspace*{2mm}
\addcontentsline{toc}{section}{Acknowledgments\,}
\section*{Acknowledgments}
\vspace*{-1.5mm}
We thank Linda M.~Carpenter and Abdelhak Djouadi for useful discussions. 
This research was supported by the NSF of China (under grants 10625522, 10635030, 11135003) 
and the National Basic Research Program of China (under grant 2010CB833000), and by Tsinghua University.



\end{document}

Title: LHC Signatures of Two-Higgs-Doublets with Fourth Family
\\
Authors: Ning Chen, Hong-Jian He,
\\
Comments: 22pp, 10 Figs, JHEP published version, references added
\\
On-going Higgs searches in the light mass window are of vital importance for testing the
Higgs mechanism and probing new physics beyond the standard model (SM). The latest ATLAS
and CMS searches for the SM Higgs boson at the LHC (7TeV) found some intriguing excesses
of events in the \gamma\gamma/VV^* channels (V=Z,W) around the mass-range of 124-126 GeV.
We explore a possible explanation of the \gamma\gamma and VV^* signals from the light
CP-odd Higgs A^0 or CP-even Higgs h^0 from the general two-Higgs-doublet model with
fourth-family fermions. We demonstrate that by including invisible decays of the
Higgs boson A^0 or h^0 to fourth-family neutrinos, the predicted \gamma\gamma and VV^*
signals can explain the observed new signatures at the LHC, and will be further probed
by the forthcoming LHC runs in 2012.